\documentclass[aps,prl,reprint,twocolumn,notitlepage,floatfix,tightenlines,superscriptaddress]{revtex4-1}
\usepackage[latin9]{inputenc}
\usepackage{natbib}
\makeatletter

\usepackage{float}
\usepackage{amsfonts}
\usepackage{amsmath}
\usepackage{amssymb}
\usepackage{amstext}
\usepackage{mathtools}
\usepackage[T1]{fontenc}
\usepackage{graphicx}

\graphicspath{{Figures/}}
\numberwithin{equation}{section}

\usepackage[colorlinks=true,citecolor=blue,linkcolor=blue,urlcolor=blue]{hyperref}

\newcommand{\even}{even}
\newcommand{\odd}{odd}

\makeatother

\begin{document}

\title{The bound-state solutions of the one-dimensional pseudoharmonic oscillator}

\author{Rufus Boyack}
\affiliation{D\'epartement de physique, Universit\'e de Montr\'eal, Montr\'eal, Qu\'ebec H3C 3J7, Canada}

\author{Asadullah Bhuiyan}
\affiliation{Department of Physics, University of Alberta, Edmonton, Alberta T6G 2E1, Canada}

\author{Aneca Su}
\affiliation{Department of Physics, University of Alberta, Edmonton, Alberta T6G 2E1, Canada}

\author{Frank Marsiglio}
\affiliation{Department of Physics, University of Alberta, Edmonton, Alberta T6G 2E1, Canada}

\begin{abstract}
We study the bound states of a quantum mechanical system consisting of a simple harmonic oscillator with an inverse square interaction, whose interaction strength is governed by a constant $\alpha$. 
The singular form of this potential has doubly-degenerate bound states for $-1/4\leq\alpha<0$ and $\alpha>0$; since the potential is symmetric, these consist of even and odd-parity states.
In addition we consider a regularized form of this potential with a constant cutoff near the origin. 
For this regularized potential, there are also even and odd-parity eigenfunctions for $\alpha\geq  -1/4$. 
For attractive potentials within the range $-1/4\leq\alpha<0$, there is an even-parity ground state with increasingly negative energy and a probability density that approaches a Dirac delta function as the cutoff parameter becomes zero. 
These properties are analogous to a similar ground state present in the regularized one-dimensional hydrogen atom. 
We solve this problem both analytically and numerically, and show how
the regularized excited states approach their unregularized counterparts. 
\end{abstract}

\maketitle

\section{Introduction}
\label{sec:Intro}

The one-dimensional (1D) potential $V\sim-a/x^{2}$ is a fascinating quantum mechanical system with several theoretical perplexities~\citep{Case1950,Gupta1993,Essin2006,Nguyen2020}, including the absence of any bound-state solutions. 
As noted in Ref.~\citep{Essin2006}, for a particle of mass $m$ in this potential there is no quantity with the dimensions of energy that can be constructed from only the
available parameters $m, \hbar$, and $a$, and thus no quantized bound-state solutions are expected to exist. 
A familiar system with a natural energy scale is the simple harmonic oscillator, where
the oscillator frequency $\omega$ determines $\hbar\omega$ as the pertinent energy scale.
In addition, the oscillator length $\sqrt{\hbar/\left(m\omega\right)}$ is the natural length scale.
Thus, one expects that, if a simple harmonic oscillator interaction is added to the $1/x^{2}$ potential, 
then bound-state solutions will exist for this combined system. 
A physical context where this situation might arise is if a harmonic oscillator is considered in
the presence of an external dipole-like interaction. 
Indeed, in Refs.~\citep{Palma2003,Palma2003b} the authors studied the 1D potential 
\begin{equation}
\label{eq:Potential1}
V\left(x\right)=\frac{1}{2}m\omega^{2}x^{2}+\frac{\hbar^{2}}{2m}\frac{\alpha}{x^{2}}
\end{equation}
in such a context, and they computed the bound-state energy eigenvalues and eigenfunctions for $\alpha>0$. 
The energy eigenfunctions were found to be doubly degenerate, 
which is in contrast to the well-known theorem~\cite{LandauLifshitz} that finite 1D potentials do not have degenerate spectra. 
For $\alpha<0$, Ref.~\cite{Palma2003} stated that the attractive potential has no lower energy bound. 
As we will find, finite-energy solutions do indeed exist when $\alpha<0$. 

Prior to the studies of Refs.~\cite{Palma2003,Palma2003b}, Ref.~\cite{Ballhausen1988} studied the potential in Eq.~\eqref{eq:Potential1}, 
and for $\alpha>0$ the same eigenfunctions and eigenvalues as given in Refs.~\cite{Palma2003,Palma2003b} were obtained. 
Interestingly, for $-1/4\leq\alpha<0$, two sets of bound-state solutions were also obtained; that is, for a fixed $\alpha$, 
two distinct bound-state eigenfunctions with distinct energy eigenvalues were found. 
This counterintuitive behaviour, namely two distinct solutions for the same value of $\alpha$, was criticized~\cite{Senn1989}, and it was argued that only one of the proposed solutions was in fact the correct one. 
This argument was validated using an alternative explanation~\cite{Ballhausen1989}, and as a result a well-defined set of bound-state solutions for $-1/4\leq\alpha<0$ was obtained. 
As we will show below, for this range of $\alpha$ a degenerate set of even and odd-parity solutions also exists. 
This is most readily seen in the regularized calculations.

The potential in Eq.~\eqref{eq:Potential1} has been studied using a variety of different methods, including raising and lowering operators~\cite{Ballhausen1988b,Singh2006,DongBook},
$su(1,1)$ spectrum generating algebra~\cite{Brajamani1990,Buyukilic1992,Levai1994,Oyewumi2012}, supersymmetric quantum mechanics~\citep{Pena2005}, Laplace transform~\cite{Arda2012}, 
and by explicitly solving the differential equation~\cite{Goldman,TerHaar,Weissman1979}; the latter approach was for fixed $\alpha=1$.
The 2D~\cite{Dong2005} and 3D~\cite{Constantinescu,Sage1984,Sage1985,Dong2003,Oyewumi2008,Tezcan2009} versions of this potential have also been studied. 
In molecular physics Eq.~\eqref{eq:Potential1} is known as a pseudoharmonic oscillator potential~\cite{Ballhausen1988} 
(strictly speaking the term pseudoharmonic oscillator usually refers to the 3D version with a specific value of $\alpha$~\cite{Oyewumi2012}). 
More detailed references on applications of pseudoharmonic-oscillator-type potentials can be found in Refs.~\citep{Oyewumi2012,Nogueira2016}, 
including the application of the 3D version to describe a diatomic molecule~\cite{Oyewumi2012}.

Singular potentials~\cite{Andrews1976} in 1D quantum mechanics are of great interest, with the most notable example being the 1D hydrogen atom. 
An essential aspect of the hydrogen problem concerns the existence of even-parity solutions, which for the singular, ``unregularized'' potential have been argued 
by different researchers to be present~\cite{Andrews1981,Andrews1981b,Home1982,Andrews1988,Hammer1988} or absent~\cite{Haines1969,Gomes1980,Gomes1981,Palma2006}, 
whereas for a regularized version of this potential~\cite{Loudon1959,Boyack2021} even-parity solutions are indisputably present.  
An interesting  phenomenon in the regularized 1D hydrogen atom is the presence of an even-parity ground state whose energy becomes increasingly negative as the cutoff parameter goes to zero~\cite{Loudon1959,Boyack2021}.
Moreover, the probability density for the corresponding wave function of this state limits to a Dirac delta function.
This ground state acts like a pseudopotential~\cite{Ibrahim2018}.

In this paper we will study a regularized form of Eq.~\eqref{eq:Potential1} and show that, for $-1/4\leq\alpha<0$, the pseudoharmonic oscillator also has an even-parity ground state 
with the exact same aforementioned properties as the ground state of the 1D hydrogen atom, namely, increasingly negative energy and a probability density limiting to a Dirac delta function. 
We will obtain the analytical form of this solution as a function of the interaction strength $\alpha$ and numerically confirm this result in the limit of small  cutoff. 
Thus, for both $-1/4\leq\alpha<0$ and $\alpha>0$, there are even and odd-parity states, and the energies of these solutions become degenerate with one another as the cutoff parameter limits to zero. 
Since there are many theoretical applications of the pseudoharmonic oscillator, as discussed previously, this analysis will be of interest in several pertinent contexts. 

The structure of the paper is as follows. In Sec.~\ref{sec:UnregSE} we review the analysis of the unregularized potential. 
We are in agreement with previous researchers, except that we argue that the eigenfunctions are also doubly degenerate for negative $\alpha$.
Following this, in Sec.~\ref{sec:RegSE1} we analyze the regularized potential for the case $-1/4\leq\alpha<0$ and demonstrate
that our results for the excited states reproduce those for the unregularized potential as the cutoff limits to zero. 
Then, in Sec.~\ref{sec:GS}, we analyze the properties of the even-parity ground state that has an increasingly negative energy as the cutoff approaches zero. 
In Sec.~\ref{sec:RegSE2}, we study the regularized potential for the case $\alpha>0$. 
In Sec.~\ref{sec:matrix_mechanics}, we describe a simple numerical method based on matrix mechanics that we have used
to confirm the analytical results obtained with the less familiar confluent hypergeometric functions.
Finally, we present our conclusions in Sec.~\ref{sec:Conclusion}. Technical details are presented in the Appendices.
In Appendix~\ref{App:OddError}, we derive the ``correction'' term for the difference between the energy eigenvalues of the regularized potential and those of the unregularized potential. 
In Appendix~\ref{App:CCoeffs}, we derive an expansion of the ground-state energy in small values of the cutoff parameter.
In Appendix~\ref{App:C0Sol}, we obtain a closed-form expression for the approximate ground-state energy as a function of $\alpha$.

\section{Unregularized potential}
\label{sec:UnregSE}

\subsection{Eigenstates and eigenvalues}

\begin{figure}[h]
\centering
\includegraphics[scale=0.4]{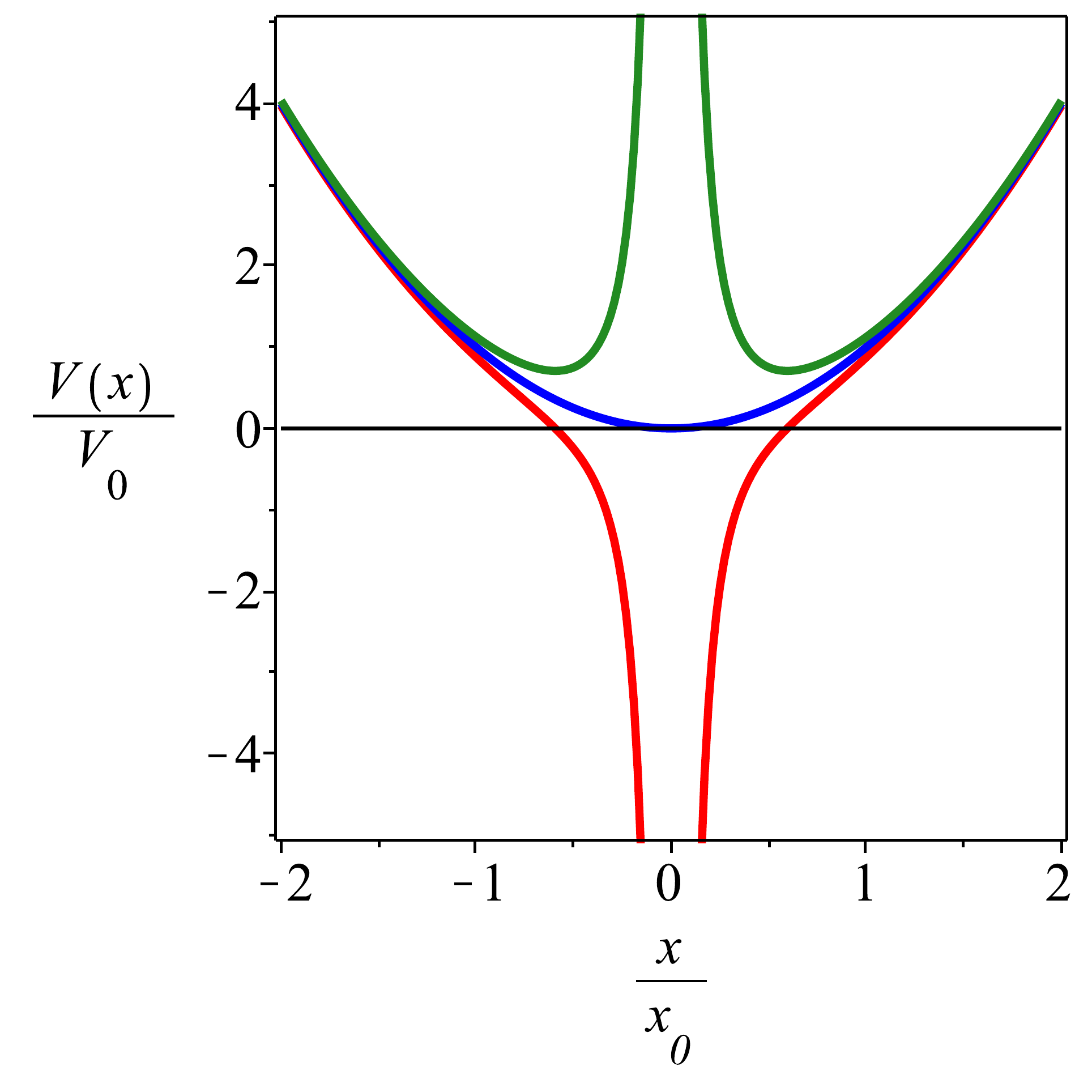}
\caption{The potential energy function in Eq.~\eqref{eq:Potential2} for the attractive case, $\alpha=-1/8$ (red), for the simple harmonic
oscillator, $\alpha=0$ (blue), and for the repulsive case, $\alpha=1/8$ (green).}
\label{fig:Potential}
\end{figure}

The potential energy can be written in a compact form by introducing the length scale $x_{0}$ and the energy scale $V_{0}$ 
defined by 
\begin{align}
\label{eq:X0} x_{0}&=\sqrt{\frac{\hbar}{m\omega}},\\
\label{eq:V0} V_{0}&=\frac{1}{2}\hbar\omega.
\end{align}
The potential is then given by
\begin{equation}
\label{eq:Potential2}
V(x)=V_{0}\left[\left(\frac{x}{x_{0}}\right)^2+\alpha\left(\frac{x_{0}}{x}\right)^2\right].
\end{equation}
A plot of the potential is shown in Fig.~\ref{fig:Potential} for various values of $\alpha$. 

The time-independent Schr$\ddot{\text{o}}$dinger equation for the
potential in Eq.~\eqref{eq:Potential1} is
\begin{equation}
-\frac{\hbar^{2}}{2m}\frac{d^{2}\psi\left(x\right)}{dx^{2}}+\left(\frac{1}{2}m\omega^{2}x^{2}+\frac{\alpha\hbar^{2}}{2m}\frac{1}{x^{2}}\right)\psi\left(x\right)=E\psi\left(x\right).\label{eq:SE1}
\end{equation}
Due to the $1/x^2$ term in the potential, we require the solutions to satisfy the boundary condition $\psi(0)=0$.
Since the Hamiltonian has inversion symmetry, the solutions of Eq.~\eqref{eq:SE1}
have definite parity and are either even or odd functions of position.
Define the dimensionless variables $y$ and $\kappa$ via
\begin{equation}
\label{eq:EandyDefs}
y = \frac{x}{x_{0}};\quad E=\left(\kappa+\frac{1}{2}\right)\hbar\omega.
\end{equation}
The Schr$\ddot{\text{o}}$dinger equation then becomes 
\begin{equation}
-\psi^{\prime\prime}\left(y\right)+\left(y^{2}+\frac{\alpha}{y^{2}}\right)\psi\left(y\right)=2\left(\kappa+\frac{1}{2}\right)\psi\left(y\right).\label{eq:SE}
\end{equation}
In the limit that $y\rightarrow\infty$, the asymptotic behaviour of $\psi(y)$ is $\psi\left(y\right)\sim e^{-\frac{1}{2}y^{2}}$. 
Therefore, we consider the ansatz $\psi\left(y\right)=e^{-\frac{1}{2}y^{2}}y^{\nu}g\left(y\right)$.
The indicial equation for $g$ motivates introducing the variable $\nu$ defined by $\alpha=\nu\left(\nu-1\right)$. 
Solving this equation gives 
\begin{equation}
\nu=\frac{1}{2}\pm\sqrt{\frac{1}{4}+\alpha}.
\end{equation}
Real solutions thus require $\frac{1}{4}+\alpha\geq0$. Let the two solutions be denoted by $\nu_{\pm}$. 
For $\alpha>0$, $\nu_{+}>1$ and $\nu_{-}<0$. 
In this case we will choose the positive root; in reality both should be considered, but the outcome will be the same~\cite{Mathews2021}.
Suffice it to say that, for $-1/4\leq\alpha<0$ and for $\alpha>0$, $\nu=\nu_{+}$ is the only physically acceptable solution. 
This requirement is consistent with the conclusions of Refs. \cite{Senn1989,Ballhausen1989} and dispels
the unphysical behaviour found in Fig.~1 of Ref.~\cite{Ballhausen1988}, which exhibited two possible energy eigenvalues for a given $\alpha$ when $-1/4\leq\alpha<0$.

After using the indicial equation, the differential equation for $g$ is
\begin{equation}
-\left[g^{\prime\prime}\left(y\right)+\frac{2\nu}{y}g^{\prime}\left(y\right)\right]+2\nu g\left(y\right)+2yg^{\prime}\left(y\right)=2\kappa g\left(y\right).
\end{equation}
Now let $w=y^{2}$. This substitution then leads to 
\begin{equation}
wg^{\prime\prime}\left(w\right)+\left(\nu+\frac{1}{2}-w\right)g^{\prime}\left(w\right)-\left(\frac{\nu-\kappa}{2}\right)g\left(w\right)=0.
\end{equation}
The confluent hypergeometric differential equation (also known as Kummer's equation) has the form 
\begin{equation}
wg^{\prime\prime}\left(w\right)+\left(b-w\right)g^{\prime}\left(w\right)-ag\left(w\right)=0
\end{equation}
and the solution is a linear combination of two independent solutions of this equation, typically taken to be $M(a,b,w)$ (known as
the Kummer function), and $U(a,b,w)$ (known as the Tricomi function). Several other possibilities exist, as recently catalogued
in Ref.~\cite{Mathews2021}. See also Refs.~\cite{AbramowitzStegun,NIST2020}. 

The result is that the solution is given by
\begin{equation}
g\left(w\right)=U\left(\frac{\nu-\kappa}{2},\nu+\frac{1}{2},w\right),
\end{equation}
where $\frac{1}{2}\left(\nu-\kappa\right)$ must be a non-positive integer, in which case the Tricomi function truncates to a polynomial. 
Therefore, $\kappa-\nu=2n$, where $n\in\mathbb{Z}_{\geq0}$. 
The relation between the generalized Laguerre polynomial and the confluent hypergeometric function is given in Eq.~(13.6.27) of Ref.~\cite{AbramowitzStegun} and Eq.~(13.6.19) of Ref.~\cite{NIST2020}:
$U\left(-n,\beta+1,w\right)=\left(-1\right)^{n}n!L_{n}^{\left(\beta\right)}\left(w\right)$, where we use the Laguerre polynomials as defined in Refs.~\cite{AbramowitzStegun,NIST2020}. 
Thus, up to a normalization constant, the solution is 
\begin{equation}
g\left(w\right)=L_{n}^{\left(\nu-\frac{1}{2}\right)}\left(w\right) = 
\sum_{s=0}^n \frac{\Gamma\left(n+\nu + \frac{1}{2}\right)}{\Gamma\left(s+\nu + \frac{1}{2}\right)}\frac{\left(-w\right)^s}{\left(n-s\right)! s!}.
\end{equation}

In summary, the complete solution for the eigenfunctions, when $y\geq0$, is
\begin{equation}
\label{eq:PsiR}
\psi_{+}\left(y\right)=A\left(-1\right)^{n}e^{-\frac{1}{2}y^{2}}y^{\nu}L_{n}^{\left(\nu-\frac{1}{2}\right)}\left(y^{2}\right),\ n\in\mathbb{Z}_{\geq0}.
\end{equation}
Similarly, the solution for $y\leq0$ is 
\begin{equation}
\label{eq:PsiL}
\psi_{-}\left(y\right)=B\left(-1\right)^{n}e^{-\frac{1}{2}y^{2}}\left(-y\right)^{\nu}L_{n}^{\left(\nu-\frac{1}{2}\right)}\left(y^{2}\right),\ n\in\mathbb{Z}_{\geq0}.
\end{equation}

\subsection{Continuity and normalization conditions}

Continuity of the wave function at the origin requires that $\psi_{+}\left(0^{+}\right)=\psi_{-}\left(0^{-}\right)$.
Since $\nu>0$, the wave function vanishes at the origin and so this condition does not impose a constraint. 
For potentials with a finite jump discontinuity, the derivative of the wave function is continuous~\cite{Branson1979,Andrews1981}.
However, since the potential in Eq.~\eqref{eq:Potential2} is singular at the origin, i.e., it has an infinite jump discontinuity, the behaviour of the derivative of the wave function is more subtle.
As pointed out in Ref.~\cite{Home1982}, excluding the case of the Dirac delta potential, the requirement of Hermiticity of the momentum operator leads to the result that a
wave function can have a discontinuous first derivative only at a point where the wave function itself vanishes. 
Both of the functions in Eqs.~\eqref{eq:PsiR}-\eqref{eq:PsiL} vanish at the origin, and as such the derivative of the wave function can be discontinuous. 
We will not impose a condition on $\psi^{\prime}$ and consider both even and odd-parity solutions.  
We now determine the normalization constant.

Let $N$ denote the normalization constant of the wave function. The normalization condition is given by
\begin{align}\label{eq:Norm}
1 & =  \int_{-\infty}^{\infty}dx\left|\psi\left(x\right)\right|^{2}\nonumber \\
 & =  N^{2}\left(\frac{\hbar}{m\omega}\right)^{\frac{1}{2}}\int_{0}^{\infty}dxx^{\nu-\frac{1}{2}}e^{-x}\left[L_{n}^{\left(\nu-\frac{1}{2}\right)}\left(x\right)\right]^{2}.
\end{align}
To evaluate the remaining integral we use the orthogonality relation for the generalized Laguerre polynomials (Eq.~(19), pg.~479 of Ref.~\cite{Prudnikov}). 
For $\lambda>-1$, we have
\begin{equation}
\int_{0}^{\infty}dxx^{\lambda}e^{-x}L_{n}^{\left(\lambda\right)}\left(x\right)L_{m}^{\left(\lambda\right)}\left(x\right)=\frac{1}{n!}\Gamma\left(n+\lambda+1\right)\delta_{n,m}.
\label{int_norm}
\end{equation}
The solution to Eq.~\eqref{eq:Norm} is thus
\begin{equation}
N=\left(\frac{m\omega}{\hbar}\right)^{\frac{1}{4}}\sqrt{\frac{n!}{\Gamma\left(n+\nu+\frac{1}{2}\right)}}.
\end{equation}
This result agrees with Eq.~(16) of Ref.~\citep{Ballhausen1988b}.

Let $\psi_{n}\left(x\right)$ be defined by 
\begin{align}
\psi_{n}\left(x\right)&=\frac{1}{\sqrt{x_{0}}}\sqrt{\frac{n!}{\Gamma\left(n+\nu+\frac{1}{2}\right)}}e^{-x^{2}/\left(2x_{0}^2\right)}\nonumber\\
&\quad\times
\left(\frac{x}{x_{0}}\right)^{\nu}L_{n}^{\left(\nu-\frac{1}{2}\right)}\left(\frac{x^2}{x_{0}^2}\right),\ x\geq0.
\end{align}
The complete solution of the problem is then given as follows. For all of the permissible (and non trivial) values of $\alpha$ for which bound-state solutions exist, $-1/4\leq\alpha<0$ and $\alpha>0$, the energy eigenvalues are
\begin{align}
\label{eq:Nu} \nu & = \frac{1}{2}+\sqrt{\frac{1}{4}+\alpha}. \\
\label{eq:En}E_{n} & = \left(2n+1+\sqrt{\frac{1}{4}+\alpha}\right)\hbar\omega,\ n\in\mathbb{Z}_{\geq0}.
\end{align}
Equations~\eqref{eq:Nu}-\eqref{eq:En} are also valid for $\alpha=0$. 
When $\alpha=0$, Eq.~\eqref{eq:Potential2} reduces to the simple harmonic oscillator potential, and another set of solutions are given by the even-parity Hermite polynomial solutions. 
Thus, for $-1/4\leq\alpha<0$ and $\alpha>0$, the energy eigenfunctions are doubly degenerate, and for $\alpha$=0 there is a ``disconnected'' set of even solutions. 
To account for this discontinuous behaviour in the energy structure of the even solutions, we label these solutions as follows. 
For $\alpha>0$, we let $n_{\even}=n_{\odd}=n\in\mathbb{Z}_{\geq0}$. However, for $-1/4\leq\alpha<0$, we let $n_{\even}-1=n_{\odd}=n\in\mathbb{Z}_{\geq0}$, 
where $n_{\even}$ and $n_{\odd}$ label the respective even and odd-parity solutions. 

\begin{figure}[t]
\centering
\includegraphics[scale=0.4]{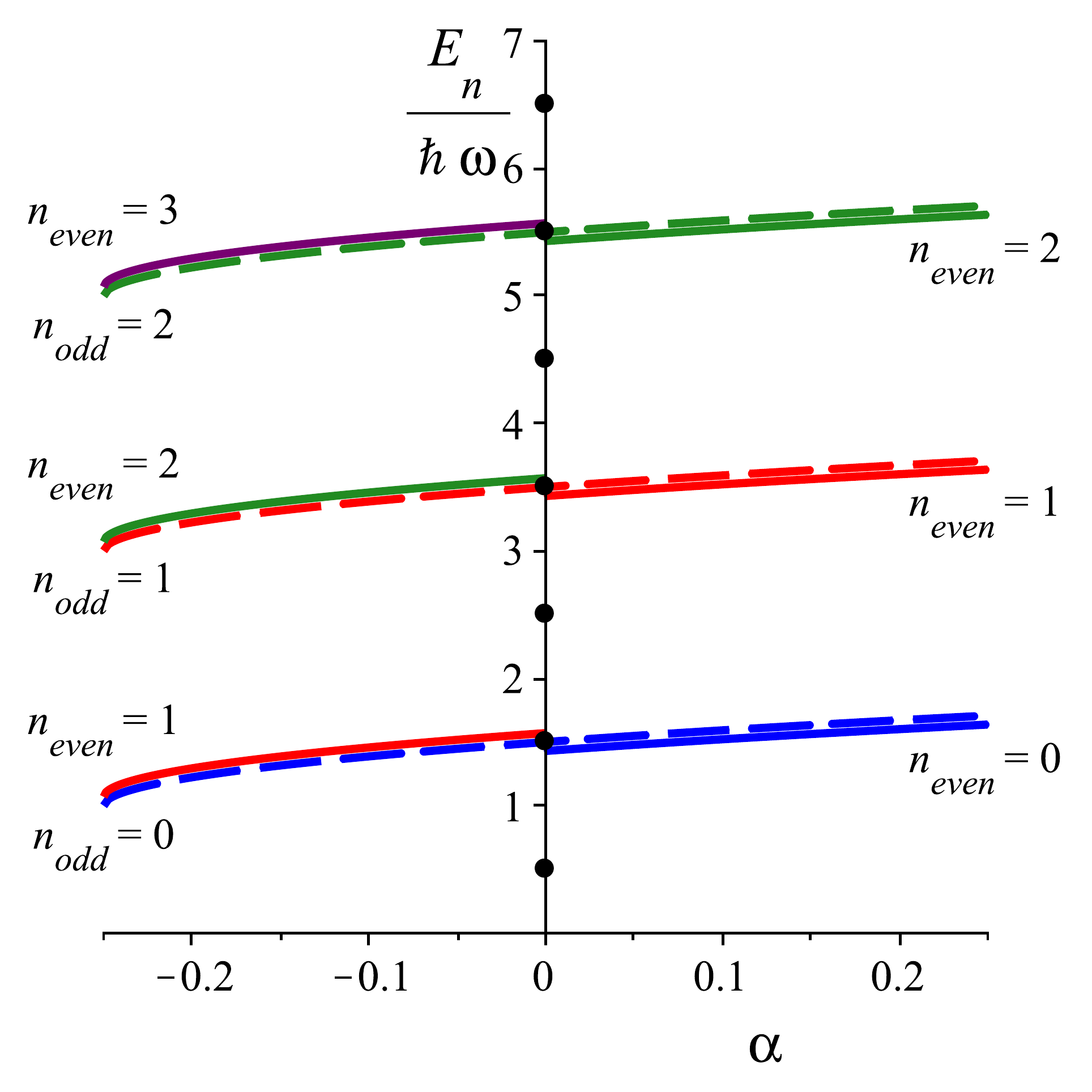}
\caption{The first few energy levels as functions of the parameter $\alpha$. We have artificially shifted the even-state eigenvalues in the positive (negative) $y$-direction for $\alpha<0\ (\alpha>0)$ for clarity. 
In reality, the even and odd-state eigenvalues lie exactly on top of each other. 
Note that all the odd-state energies (denoted by dashed coloured curves) are continuous and pass through the well-known result for the harmonic oscillator at $\alpha = 0$. 
In contrast, the even states (denoted by coloured solid curves) have a discontinuity at $\alpha = 0$; 
the actual energy values for $\alpha = 0$ lie halfway between this discontinuity. 
Moreover, the lowest even state ($n_{even} = 0$, shown in solid blue) exists only for $\alpha > 0$. 
Where is it for $\alpha < 0$? Insight into the answer to this question will be provided when we consider the regularized potential.} 
\label{fig:UnregEigs}
\end{figure}

A plot of these energy eigenvalues, as functions of $\alpha$, is shown in Fig.~\ref{fig:UnregEigs}. The solid (dashed) lines correspond to the even (odd) solutions.
As illustrated, the energy eigenvalues are doubly degenerate for $-1/4\leq\alpha<0$ and $\alpha>0$. 
A possibly counter-intuitive feature of this plot is our choice of labelling for the even solutions. 
As $\alpha$ changes sign, for a fixed $n_{\odd}$, the even solution that is degenerate with the odd solution has a different $n_{\even}$ label. 
The actual mathematical expression for the even-parity wave functions does not change as $\alpha$ changes sign; that is, the $n$ in Eqs.~\eqref{eq:PsiR} and \eqref{eq:PsiL} 
is always the same as $\alpha$ changes sign, it is merely that we define an $n_{\even}$ that is different for positive and negative $\alpha$. 
The motivation for this choice will be clearer in the next section when we study the regularized potential. 
Finally, note that all of the eigenvalues are positive, i.e., there is no negative-energy state that takes advantage of the negative potential. 
The presence of such a state will be shown for the case of the regularized potential.

\begin{figure}[t]
\centering
\includegraphics[scale=0.4]{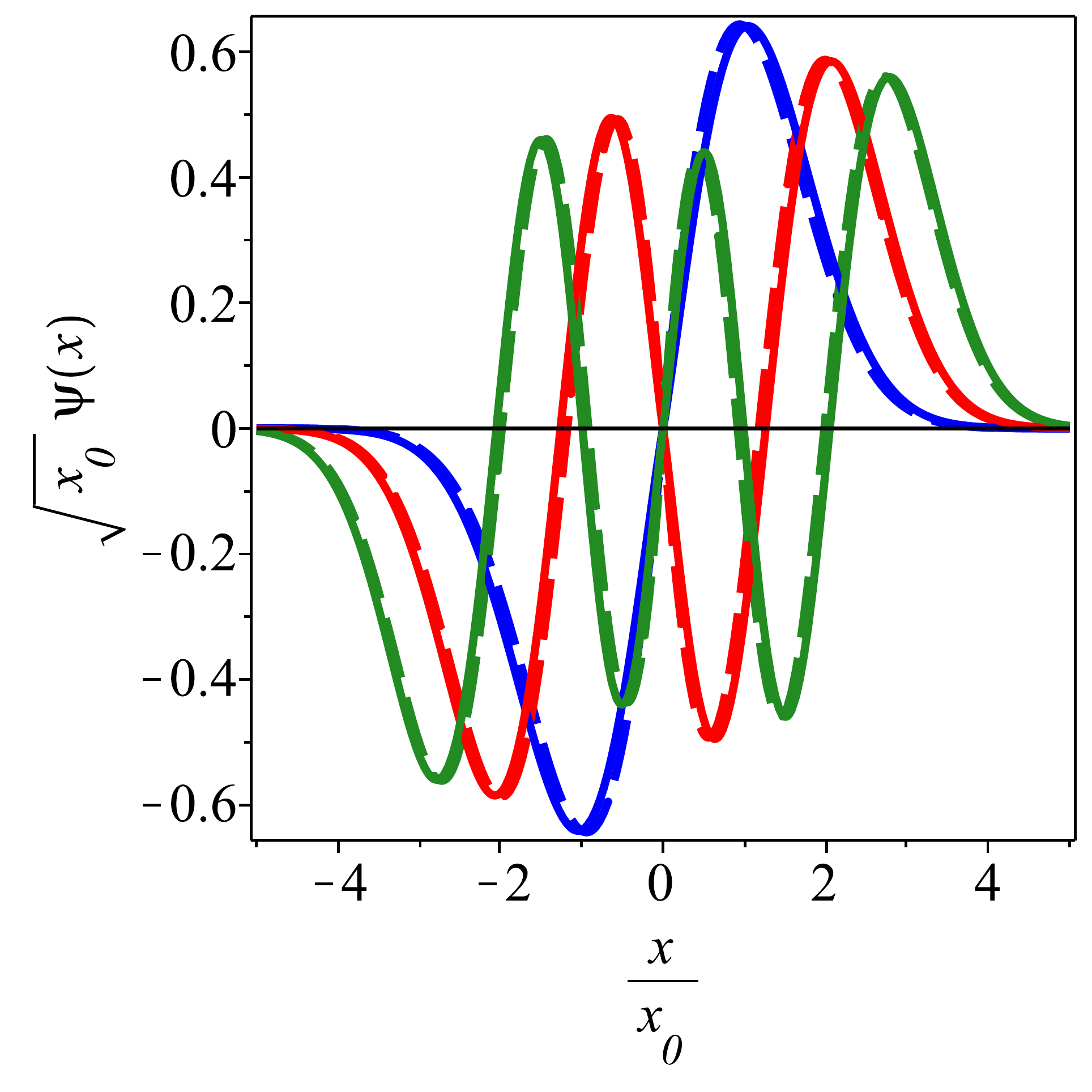}
\caption{The $n_{\odd}=0$ (blue), $n_{\odd}=1$ (red), and $n_{\odd}=2$ (green) odd-parity wave functions. The dashed curves correspond to $\alpha=-0.1$ while the solid curves are for $\alpha=0.1$.}
\label{fig:UnregPsiOdd}
\end{figure}

\begin{figure}[t]
\centering
\includegraphics[scale=0.4]{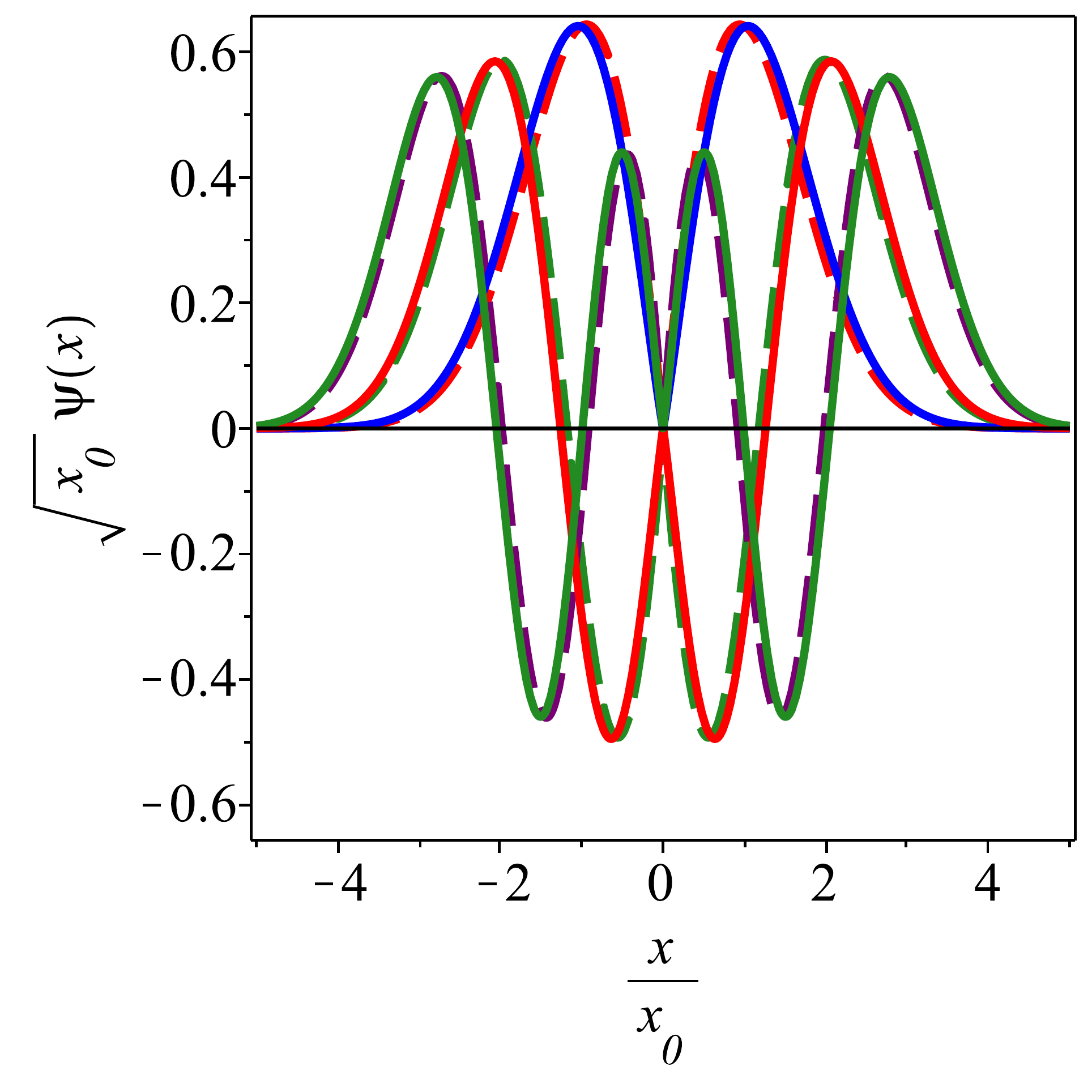}
\caption{The even-parity wave functions. The dashed curves correspond to $\alpha=-0.1$ while the solid curves are for $\alpha=0.1$. For positive $\alpha$, the colour scheme is $n_{\even}=0$ (blue), $n_{\even}=1$ (red), and $n_{\even}=2$ (green), 
whereas for negative $\alpha$, $n_{\even}=1$ (red), $n_{\even}=2$ (green), and $n_{\even}=3$ (purple).}
\label{fig:UnregPsiEven}
\end{figure}

The eigenfunctions are degenerate, and we select the even and odd-parity combinations:
\begin{equation}
\alpha\geq-1/4:\quad  \psi\left(x\right) =
\begin{dcases}
 \ \ \psi_{n}\left(x\right),\ \ \ x\geq0\\
 \pm\psi_{n}\left(-x\right),\ x\leq0.
\end{dcases}
\label{eq:Wavefn}
\end{equation}
In Figs.~\ref{fig:UnregPsiOdd} and \ref{fig:UnregPsiEven} we show the first few even and odd-parity wave functions, respectively, for positive and negative values of $\alpha$. 

\subsection{Discussion}

The energy eigenvalues obtained here agree with those derived in Ref.~\citep{Ballhausen1988}.
Importantly, there are bound-state solutions for negative values of $\alpha$ in the range $-1/4\leq\alpha<0$. 
The existence of bound-state solutions in this regime is not too surprising, since a harmonic oscillator potential encloses the singular potential; 
moreover, these bound states all have positive energy, consistent with the fact that there
are no bound states for the $1/x^{2}$ potential in this regime~\citep{Essin2006}, as this would require a negative-energy solution.

For $-1/4\leq\alpha<0$, we have argued that there are two degenerate solutions (as there are for $\alpha >0$).
The critical value $\alpha=-1/4$ is analogous to the critical field for which the fall of a particle to the centre of the potential becomes possible: see Sec.~35 of Ref.~\cite{LandauLifshitz}. 
Note that a remarkable discontinuity occurs at $\alpha=0$, where the states are no longer degenerate and the eigenvalues form the familiar ladder series (see Fig.~\ref{fig:UnregEigs}).

An interesting aspect of this problem is the double degeneracy of the bound-state solutions. 
A standard theorem~\cite{LandauLifshitz} in quantum mechanics in one dimension asserts that for finite potentials the bound-state wave functions are non degenerate.
However, for singular potentials, this theorem is modified~\cite{Andrews1976}.  

Here we have derived the exact energy eigenvalues and eigenfunctions. 
If these solutions were not already known, it would be natural to consider the $\alpha/x^2$ potential as a perturbation to a simple harmonic oscillator system 
and use non-degenerate perturbation theory to obtain the corrected eigenvalues and eigenfunctions in powers of $\alpha$.
The energy obtained~\cite{Aguilera1991} to second order in perturbation theory agrees with the expansion of the exact energy.
However, the perturbed wave functions disagree with the expansion of the exact wave functions. 
Indeed, the expansion of $x^{\nu}$ for small $\alpha$ produces a logarithmic term in $x$, which cannot arise from the sum of unperturbed eigenfunctions consisting of Hermite polynomials.
Thus, perturbation theory for this potential is singular; see Ref.~\cite{Aguilera1991} for further discussion of these points. 

Note that both the odd-parity (Fig.~\ref{fig:UnregPsiOdd}) and the even-parity (Fig.~\ref{fig:UnregPsiEven}) wave functions
are essentially identical for positive $\alpha$ and negative $\alpha$. 
Of course, given Eq.~(\ref{eq:Wavefn}), the even and odd wave functions are identical to one another for a given value of $\alpha>0$ as well. 
The significance of the first statement, however, is profound. 
This equivalence means that the singular, attractive well ($\alpha < 0$) acts as a barrier in very much the same way as the repulsive barrier ($\alpha > 0$) does.
For a simpler model this was readily understood as a consequence of the so-called pseudopotential effect \cite{Ibrahim2018}.
This effect is summarized by the following: the existence of a very negative energy bound state serves to act as a 
pseudopotential for other, higher energy states because these states must be orthogonal to the negative-energy bound state.
Because the very negative energy bound-state wave function will be strongly peaked near the origin, it will serve as an effective
barrier with respect to tunnelling in the positive-energy states.
In our case, however, we have been unable 
to identify such a negative-energy bound state.
We speculate that nonetheless it is present, but outside of the Hilbert space that we have explored.

Additional evidence comes from the cusp that is clearly present at the origin in the even-parity wave functions depicted in 
Fig.~\ref{fig:UnregPsiEven}. It is easy to show that the second derivative of this cusp-like feature produces a (repulsive) Dirac delta function.
As there is no Dirac delta function in the potential we are studying, we understand this inferred $\delta$-function to be the result
of a bound state not contained within our Hilbert space. This interpretation of our results for the unregularized potential is further
supported by results of the regularized potential.

We now investigate a regularized version of Eq.~\eqref{eq:Potential2} and study the interesting properties that arise in the limit that the cutoff is taken to zero. 
We will recover the bound states of the unregularized potential, but, in addition, a new, negative energy bound state arises, and
plays a role in causing the degeneracy in the positive-energy solutions of the regularized potential. We believe this ground state is
the one inferred above in the unregularized theory.

\section{Regularized potential case i: $-1/4\leq\alpha<0$ }
\label{sec:RegSE1}

We consider the potential  
\begin{equation}
V\left(x\right)=\left\{ \begin{array}{c}
V_{0}\left[\left(\frac{x}{x_{0}}\right)^{2}+\alpha\left(\frac{x_{0}}{x}\right)^{2}\right],\ x\geq\delta x_{0}\\ \\
V_{0}\left(\delta^{2}+\frac{\alpha}{\delta^{2}}\right), \hspace{15mm}\ x\leq\delta x_{0}.
\end{array}\right.
\label{regularized_potential}
\end{equation}
The parameter $x_{0}$ is the  oscillator length defined in Eq.~\eqref{eq:X0}, $V_{0}$ is the energy scale defined in Eq.~\eqref{eq:V0},
and $0<\delta\ll1$ is a fixed cutoff parameter used to ``regularize'' the singularity at the origin. 
We define $\widetilde{V}_{0}=V_{0}\left(\delta^{2}+\frac{\alpha}{\delta^{2}}\right)$.
In this section we consider the case $-1/4\leq\alpha<0$. 
For bound-state solutions of energy $E$ we require $E<V\left(\pm\infty\right)=\infty$. 
In addition, a theorem~\cite{LandauLifshitz} of one-dimensional quantum mechanics is that $E>V_{\text{min}}=\widetilde{V}_{0}$. 
Since $\widetilde{V}_{0}\rightarrow-\infty$ as $\delta\rightarrow0$, for the case $-1/4\leq\alpha<0$, the allowed bound-state energies in this limit are $-\infty<E<\infty$. 
In Sec.~\ref{sec:GS}, we shall show that there is indeed a ground-state solution with increasingly negative energy; that is, $E\rightarrow-\infty$ as $\delta\rightarrow0$. 
All of the other bound states correspond to excited states that have positive energy.

\subsection{Odd-parity solutions}

To derive the even and odd-parity eigenfunctions, it suffices to consider only $x\geq0$. 
We then divide space into region I: $x\leq\delta x_{0}$ and region II: $x\geq\delta x_{0}$. 
We define $q^{2}$ by $q^{2}=\frac{2m}{\hbar^{2}}\left(E-\widetilde{V}_{0}\right)$, where $E$ is defined in Eq.~\eqref{eq:EandyDefs}.
Using the definitions of $q$ and $E$, the quantity $q\delta x_{0}$ can be expressed in terms of $\kappa$ as follows:
\begin{equation}
\label{eq:QX0}
\left(q\delta x_{0}\right)^{2}=\left(2\kappa+1\right)\delta^{2}-\left(\delta^{4}+\alpha\right).
\end{equation}
In region I, the Schr$\ddot{\text{o}}$dinger equation is 
\begin{equation}
\label{eq:SERegionI_Eq}
\psi^{\prime\prime}\left(x\right)+q^{2}\psi\left(x\right)=0.
\end{equation}
The solution is 
\begin{equation}
\label{eq:SERegionI_Sol}
\psi_{\text{I}}\left(x\right)=A_{\text{I}}\cos\left(qx\right)+B_{\text{I}}\sin\left(qx\right).
\end{equation}
The solutions have either even or odd parity. Let us first consider the odd-parity solutions: $A_{\text{I}}=0$. 
In region II the Schr$\ddot{\text{o}}$dinger equation is 
\begin{equation}
\label{eq:SERegionII_Eq}
-\frac{\hbar^{2}}{2m}\psi^{\prime\prime}\left(x\right)+V_{0}\left[\left(\frac{x}{x_{0}}\right)^{2}+\alpha\left(\frac{x_{0}}{x}\right)^{2}\right]\psi\left(x\right)=E\psi\left(x\right).
\end{equation}
Following the analysis performed in Sec.~\ref{sec:UnregSE}, where $y$ and $E$ are as given in Eq.~\eqref{eq:EandyDefs}, and $\nu$ is defined as in Eq.~\eqref{eq:Nu}, the solution to this differential equation is 
\begin{align}
\label{eq:SERegionII_Sol}
\psi_{\text{II}}\left(x\right)&=A_{\text{II}}M\left(\frac{\nu-\kappa}{2},\nu+\frac{1}{2},y^{2}\right)\nonumber\\
&\quad+B_{\text{II}}U\left(\frac{\nu-\kappa}{2},\nu+\frac{1}{2},y^{2}\right).
\end{align}
For non-singular behaviour as $x\rightarrow\infty$, we require $A_{\text{II}}=0$.
The general, odd-parity solution (for $x\geq0$) is then 
\begin{equation}
\psi\left(x\right)=\left\{ \begin{array}{c}
\qquad  \qquad B_{\text{I}}\sin\left(qx\right),\hspace{18mm}  x\leq\delta x_{0}\\
B_{\text{II}}y^{\nu}e^{-\frac{1}{2}y^2}U\left(\frac{\nu-\kappa}{2},\nu+\frac{1}{2},y^2\right),\ \ \ x\geq\delta x_{0}.
\end{array}\right.\label{eq:PsiOdd}
\end{equation}

\begin{figure}[t]
\centering
\includegraphics[scale=0.4]{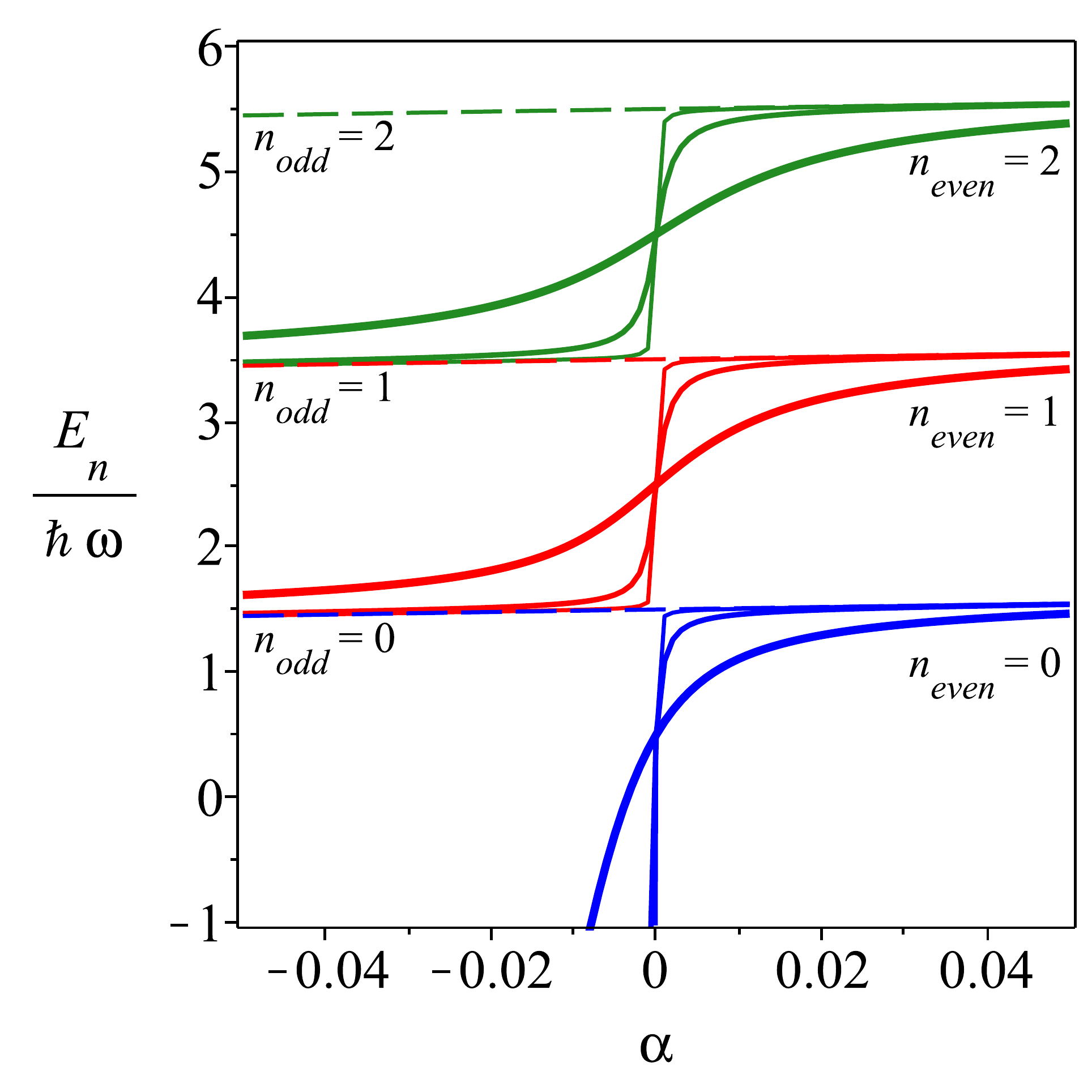}
\caption{Energy eigenvalues for the regularized pseudoharmonic oscillator potential. The odd-parity solutions correspond to the dashed lines while the even-parity solutions appear as solid lines. Three values of $\delta$ are used: $\delta=0.01, 0.001,0.0001$.
The thick line has $\delta=0.01$, the medium-sized line has $\delta=0.001$, and the thin line has $\delta=0.0001$. All of the odd-parity solutions appear approximately on the same curve, and are therefore insensitive to both the value of $\alpha$ and the value of
the regularization parameter $\delta$ over the ranges shown.} 
\label{fig:RegEigs}
\end{figure}

The energy eigenvalues are determined from the continuity of $\psi^{\prime}/\psi$ at $x=\delta x_{0}$. 
In region I, we have 
\begin{equation}
\label{eq:DPsiRegI}
\frac{\psi_{\text{I}}^{\prime}}{\psi_{\text{I}}}=q\cot\left(q\delta x_{0}\right).
\end{equation}
In region II we have 
\begin{align}
\frac{\psi_{\text{II}}^{\prime}}{\psi_{\text{II}}} & = \frac{1}{x_{0}}\left[\frac{\nu}{\delta}-\delta+2\delta\frac{U^{\prime}\left(\frac{\nu-\kappa}{2},\nu+\frac{1}{2},\delta^{2}\right)}{U\left(\frac{\nu-\kappa}{2},\nu+\frac{1}{2},\delta^{2}\right)}\right]\nonumber \\
 & = \frac{1}{x_{0}}\left[\frac{\nu}{\delta}-\delta-\delta\left(\nu-\kappa\right)\frac{U\left(\frac{\nu-\kappa}{2}+1,\nu+\frac{3}{2},\delta^{2}\right)}{U\left(\frac{\nu-\kappa}{2},\nu+\frac{1}{2},\delta^{2}\right)}\right].
\end{align}
In the last step we used Eq.~(13.4.21) of Ref.~\cite{AbramowitzStegun}: $U^{\prime}\left(a,b,z\right)=-aU\left(a+1,b+1,z\right)$.
The Tricomi function $U\left(a,b,z\right)$ obeys the following recurrence relations (see Eqs.~(13.4.17)-(13.4.18) of Ref.~\cite{AbramowitzStegun}):
\begin{align}
\label{eq:ASIdentity1}0&=U\left(a,b,z\right)-aU\left(a+1,b,z\right)-U\left(a,b-1,z\right).\\
\label{eq:ASIdentity2}0&=\left(b-a\right)U\left(a,b,z\right)+U\left(a-1,b,z\right)-zU\left(a,b+1,z\right).
\end{align}
Using these identities, we then have 
\begin{equation}
\frac{\psi_{\text{II}}^{\prime}}{\psi_{\text{II}}}=\frac{1}{\delta x_{0}}\left[\delta^{2}-\kappa-1-2\frac{U\left(\frac{\nu-\kappa}{2}-1,\nu+\frac{1}{2},\delta^{2}\right)}{U\left(\frac{\nu-\kappa}{2},\nu+\frac{1}{2},\delta^{2}\right)}\right].
\end{equation}
Equating this expression with Eq.~\eqref{eq:DPsiRegI}, we then obtain the eigenvalue condition for odd-parity solutions: 
\begin{equation}
\label{eq:EVCOdd}
q\delta x_{0}\cot\left(q\delta x_{0}\right)=\delta^{2}-\kappa-1-2\frac{U\left(\frac{\nu-\kappa}{2}-1,\nu+\frac{1}{2},\delta^{2}\right)}{U\left(\frac{\nu-\kappa}{2},\nu+\frac{1}{2},\delta^{2}\right)}.
\end{equation}
The quantity $q\delta x_{0}$ is given in Eq.~\eqref{eq:QX0}, therefore Eq.~\eqref{eq:EVCOdd} can be used to determine $\kappa$, and thus $E$, for given values of $\alpha$ and $\delta$. 
A plot of the energy eigenvalues, for negative and positive values of $\alpha$ and for even and odd-parity states, is shown in Fig.~\ref{fig:RegEigs}. 
The analytical consideration of the even-parity solutions in the case of negative $\alpha$, and also the even and odd-parity solutions for positive $\alpha$, will be considered in the subsequent sections, 
but here we present the complete results for convenience. 

One striking feature observable in Fig.~\ref{fig:RegEigs}, in the case of negative $\alpha$, is the presence of an even solution with energy that is becoming increasingly negative as $\delta\rightarrow0$. The three curves for each energy level correspond to $\delta = 0.01,
0.001, 0.0001$, with the smallest value given by the thinnest curve; these have the steepest slopes near $\alpha = 0$. 
For the lowest-energy solution (blue curves) these very small $\delta$ results are practically vertical near $\alpha = 0$.  
This solution is absent in Fig.~\ref{fig:UnregEigs} for the eigenvalues of the unregularized problem. 
The analytical properties of this state will be analyzed in more detail in Sec.~\ref{sec:GS}. 
In particular, in Sec.~\ref{sec:GSEnergy} it will be shown that the energy for this state goes as $E\sim-1/\delta^2+O\left(\delta^2\right)$, as $\delta\rightarrow0$. 
In addition, in Sec.~\ref{sec:GSPsi} it will be shown that the probability density for this state limits to a Dirac delta function. 
A similar bound-state solution is also present in the regularized 1D hydrogen atom~\cite{Boyack2021}. 

Another interesting feature in Fig.~\ref{fig:RegEigs} is that the curves are continuous functions of $\alpha$. That is, for both even and odd solutions, as $\alpha$ changes sign the energy levels vary smoothly. 
This figure should be contrasted with Fig.~\ref{fig:UnregEigs}, where the even solutions have a seemingly discontinuous behaviour as $\alpha$ changes sign. 
For example, in Fig.~\ref{fig:UnregEigs}, when $\alpha<0$ the $n_{\even}=1$ solution is degenerate with the $n_{\odd}=0$ solution, whereas when $\alpha>0$ the $n_{\even}=1$ solution is degenerate with the $n_{\odd}=1$ solution.
This behaviour can now be understood as the $\delta\rightarrow0$ limit of Fig.~\ref{fig:RegEigs}, where this crossover feature emerges naturally. 

The next step is to take the limit $\delta\rightarrow0$.
From Eq.~\eqref{eq:QX0}, we obtain $\left(q\delta x_{0}\right)^{2}\rightarrow-\alpha=\left|\alpha\right|$ as $\delta\rightarrow0$. 
Thus, $q\delta x_{0}\cot\left(q\delta x_{0}\right)\rightarrow\sqrt{\left|\alpha\right|}\cot\sqrt{\left|\alpha\right|}.$
In the previous section we found that for the unregularized potential the parameter $\kappa$ is given by $\kappa=2n+\nu$. 
Based on this result, for the regularized potential we then define
\begin{equation}
\label{eq:KEqn}
\kappa=2n+\nu+2\epsilon_{n},\ n\in\mathbb{Z}_{\geq0}.
\end{equation}
The ``correction term'' $\epsilon_{n}$ characterizes the difference between the energy eigenvalues for the unregularized and regularized potentials. 
For the unregularized potential, $\epsilon_{n}=0$. The next goal is to determine $\epsilon_{n}$ as a function of $\alpha$ in the limit $\delta\ll1$. 
As shown in Appendix~\ref{App:OddError}, the expression for $\epsilon_{n}$ in the limit $\delta\ll1$ is
\begin{align}
\label{eq:OddError1}
\epsilon_{n}&=\left(\frac{\nu-\sqrt{\left|\alpha\right|}\cot\sqrt{\left|\alpha\right|}}{\nu-1+\sqrt{\left|\alpha\right|}\cot\sqrt{\left|\alpha\right|}}\right)\nonumber\\
&\quad\times\frac{\left(-1\right)^{n}}{\Gamma\left(\frac{1}{2}-\nu-n\right)n!}\frac{\Gamma\left(\frac{3}{2}-\nu\right)}{\Gamma\left(\nu+\frac{1}{2}\right)}\delta^{2\nu-1}.
\end{align}
Since $\nu>1/2$, $\epsilon_{n}\rightarrow0$ as $\delta\rightarrow0$. 

\begin{figure}[t]
\centering
\includegraphics[scale=0.4]{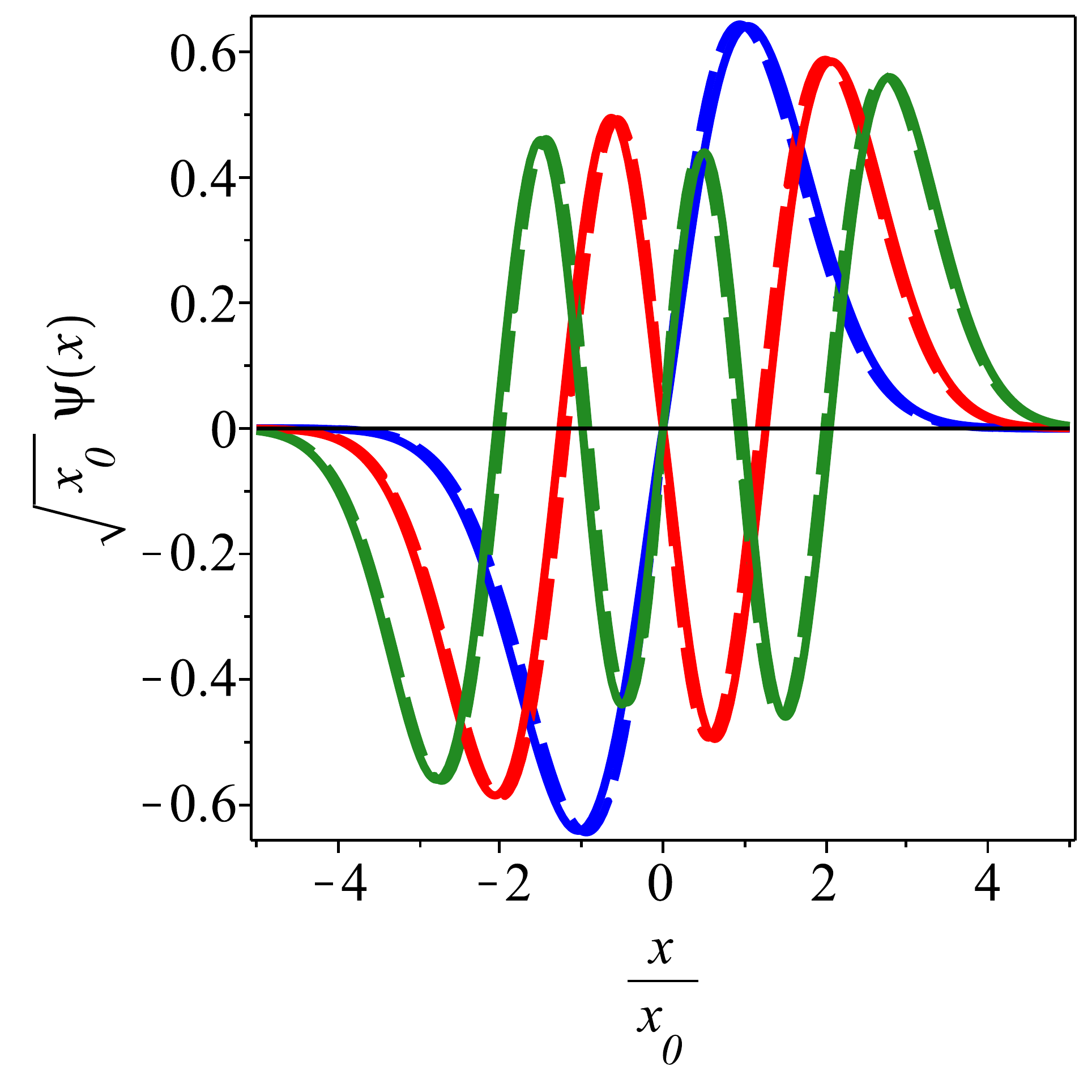}
\caption{The $n_{\odd}=0$ (blue), $n_{\odd}=1$ (red), and $n_{\odd}=2$ (green) odd-parity wave functions for the regularized potential with $\delta=0.01$. The dashed curves correspond to $\alpha=-0.1$ while the solid curves are for $\alpha=0.1$. A comparison with
the results of Fig.~\ref{fig:UnregPsiOdd} shows that, even for this relatively large value of the regularization parameter, $\delta = 0.01$, the
regularized and unregularized results are almost identical. For even smaller values of $\delta$ (not shown) the two sets of curves become
identical.}
\label{fig:RegPsiOdd}
\end{figure}

Let us now turn to the wave function for odd-parity states, given in Eq.~\eqref{eq:PsiOdd}.
Continuity of $\psi$ at $x=\delta x_{0}$ imposes the condition
\begin{equation}
\label{eq:BIIOdd}
B_{\text{I}}\sin\left(q\delta x_{0}\right)=B_{\text{II}}\delta^{\nu}e^{-\delta^{2}/2}U\left(\frac{\nu-\kappa}{2},\nu+\frac{1}{2},\delta^{2}\right).
\end{equation}
Normalization of $\psi$ requires $\int_{-\infty}^{\infty}dx\left|\psi\left(x\right)\right|^{2}=1$. 
After inserting Eq.~\eqref{eq:BIIOdd} into Eq.~\eqref{eq:PsiOdd}, then performing the normalization integral and solving for $B^{2}_{\text{I}}$, we obtain
\begin{align}
\label{eq:BIOdd}
B^{2}_{\text{I}}&=\frac{1}{2\delta x_{0}}\Biggl\{ \int_{0}^{1}dy\sin^{2}\left(q\delta x_{0}y\right)+\sin^{2}\left(q\delta x_{0}\right)\int_{1}^{\infty}dyy^{2\nu}\nonumber\\
&\quad\times e^{-\delta^{2}\left(y^{2}-1\right)}\left[\frac{U\left(\frac{\nu-\kappa}{2},\nu+\frac{1}{2},\delta^{2}y^{2}\right)}{U\left(\frac{\nu-\kappa}{2},\nu+\frac{1}{2},\delta^{2}\right)}\right]^{2}\Biggr\} ^{-1}.
\end{align}
The first few odd-parity wave functions are shown in Fig.~\ref{fig:RegPsiOdd}. For completeness, we present the results for positive and negative $\alpha$. The analysis for the positive $\alpha$ case is deferred to Sec.~\ref{sec:RegSE2OddPosAlpha}.
These wave functions strongly resemble the results for the unregularized potential shown in Fig.~\ref{fig:UnregPsiOdd}. 
In fact, for even smaller values of $\delta$ (not shown), these curves become indistinguishable from those of Fig.~\ref{fig:UnregPsiOdd}.
In the next section we investigate the even-parity solutions. 

\subsection{Even-parity solutions}

\begin{figure}[t]
\centering
\includegraphics[scale=0.4]{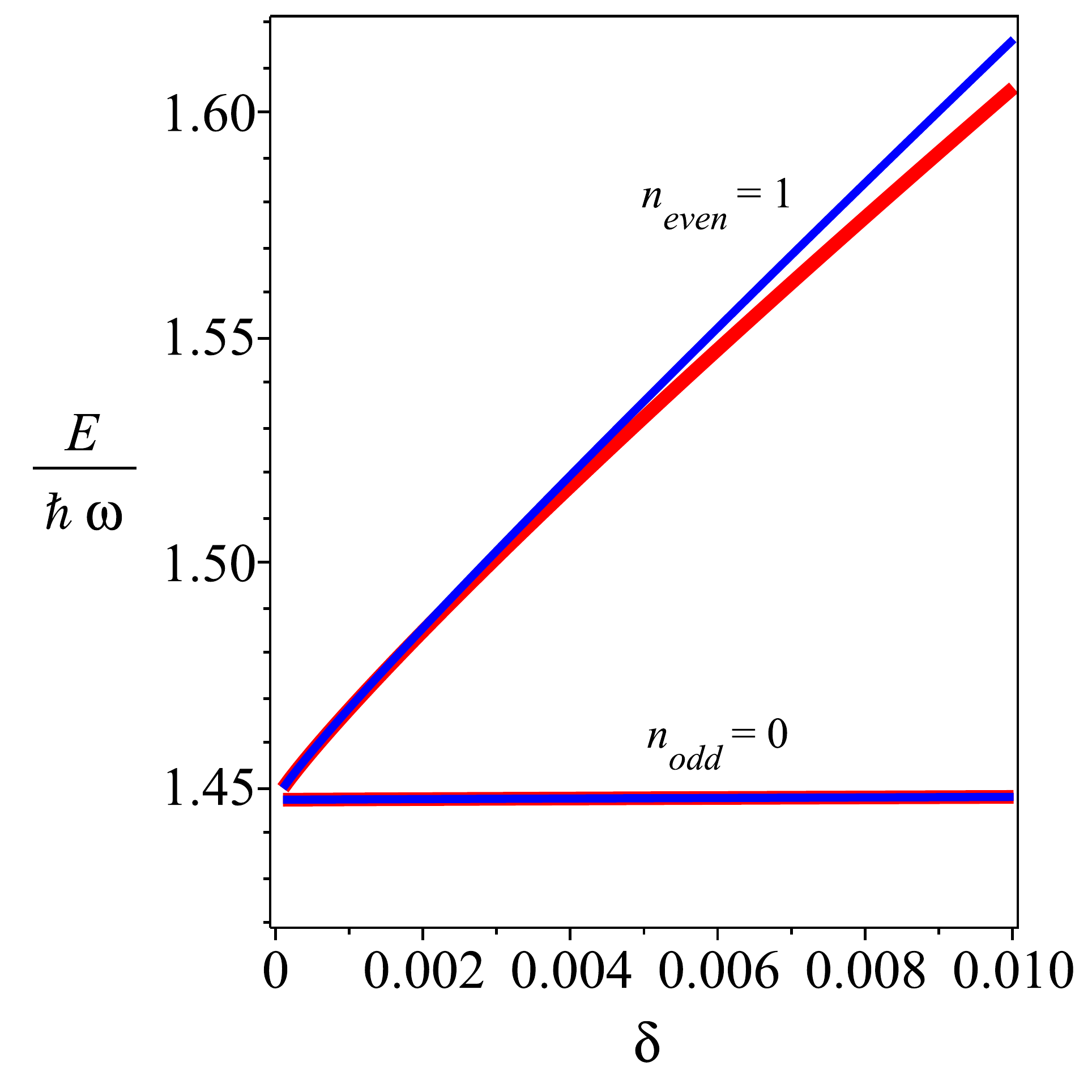}
\caption{Exact energy eigenvalues for the even and odd solutions (blue) versus the approximate energy eigenvalues (red). 
We have made the red curve thicker so that it is visible, since the energies are extremely close, particularly in the case of the odd solutions. 
Here $\alpha=-0.05$.}
\label{fig:EnergyCorrNegAlpha}
\end{figure}

The Schr$\ddot{\text{o}}$dinger equation in region I is given in Eq.~\eqref{eq:SERegionI_Eq}, with the general solution given in Eq.~\eqref{eq:SERegionI_Sol}. For even-parity solutions we set $B_{\text{I}}=0$. 
The general solution for the Schr$\ddot{\text{o}}$dinger equation in region II is given in Eq.~\eqref{eq:SERegionII_Sol},  where we set $A_{\text{II}}=0$. 
For even-parity solutions, the wave function is given by
\begin{equation}
\psi\left(x\right)=\left\{ \begin{array}{c}
\qquad \qquad A_{\text{I}}\cos\left(qx\right),\hspace{15mm}  x\leq\delta x_{0}\\
B_{\text{II}}y^{\nu}e^{-\frac{1}{2}y^2}U\left(\frac{\nu-\kappa}{2},\nu+\frac{1}{2},y^2\right),\ x\geq\delta x_{0}.
\end{array}\right.\label{eq:PsiEven}
\end{equation}
The eigenvalue condition is again determined by requiring continuity of $\psi^{\prime}/\psi$ at $x=\delta x_{0}$. In contrast to Eq.~\eqref{eq:EVCOdd} for odd-parity solutions, 
the result for even-parity solutions is given by
\begin{equation}
\label{eq:EVCEven}
q\delta x_{0}\tan\left(q\delta x_{0}\right)=\kappa+1-\delta^{2}+2\frac{U\left(\frac{\nu-\kappa}{2}-1,\nu+\frac{1}{2},\delta^{2}\right)}{U\left(\frac{\nu-\kappa}{2},\nu+\frac{1}{2},\delta^{2}\right)}.
\end{equation}
Following the analysis in Appendix~\ref{App:OddError}, we can determine the correction term in a manner similar
to that used for the odd states. The only difference in this case is the replacement of $\cot$ by $-\tan$ in Eq.~\eqref{eq:OddError1}.
Thus, the expression for $\epsilon_{n}$ for the even-parity states is
\begin{align}
\label{eq:EvenError1}
\epsilon_{n}&=\left(\frac{\nu+\sqrt{\left|\alpha\right|}\tan\sqrt{\left|\alpha\right|}}{\nu-1-\sqrt{\left|\alpha\right|}\tan\sqrt{\left|\alpha\right|}}\right)\nonumber\\
&\quad\times\frac{\left(-1\right)^{n}}{\Gamma\left(\frac{1}{2}-\nu-n\right)n!}\frac{\Gamma\left(\frac{3}{2}-\nu\right)}{\Gamma\left(\nu+\frac{1}{2}\right)}\delta^{2\nu-1}.
\end{align}
Note that, for negative $\alpha$, in the above formula we replace $n_{\even}-1=n$, as illustrated in Fig.~\ref{fig:RegEigs}, where $n_{\even}$ starts from 1,2,3,\dots. 
This redefinition merely amounts to a relabelling. 
In Fig.~\ref{fig:EnergyCorrNegAlpha}, we compare the exact energy eigenvalues (shown in blue), computed using Eqs.~\eqref{eq:EVCOdd} and \eqref{eq:EVCEven} for odd and even solutions, respectively, 
against those determined using Eqs.~\eqref{eq:KEqn}, \eqref{eq:OddError1}, and \eqref{eq:EvenError1} (shown in red) for small values of $\delta$ and $\alpha=-0.05$. 
The results are in very good agreement for small values of $\delta$. The relative agreement for the odd corrections is even better, as
a zoom of Fig.~\ref{fig:EnergyCorrNegAlpha} focussing only on the odd correction indicates (not shown).

\begin{figure}[t]
\centering
\includegraphics[scale=0.4]{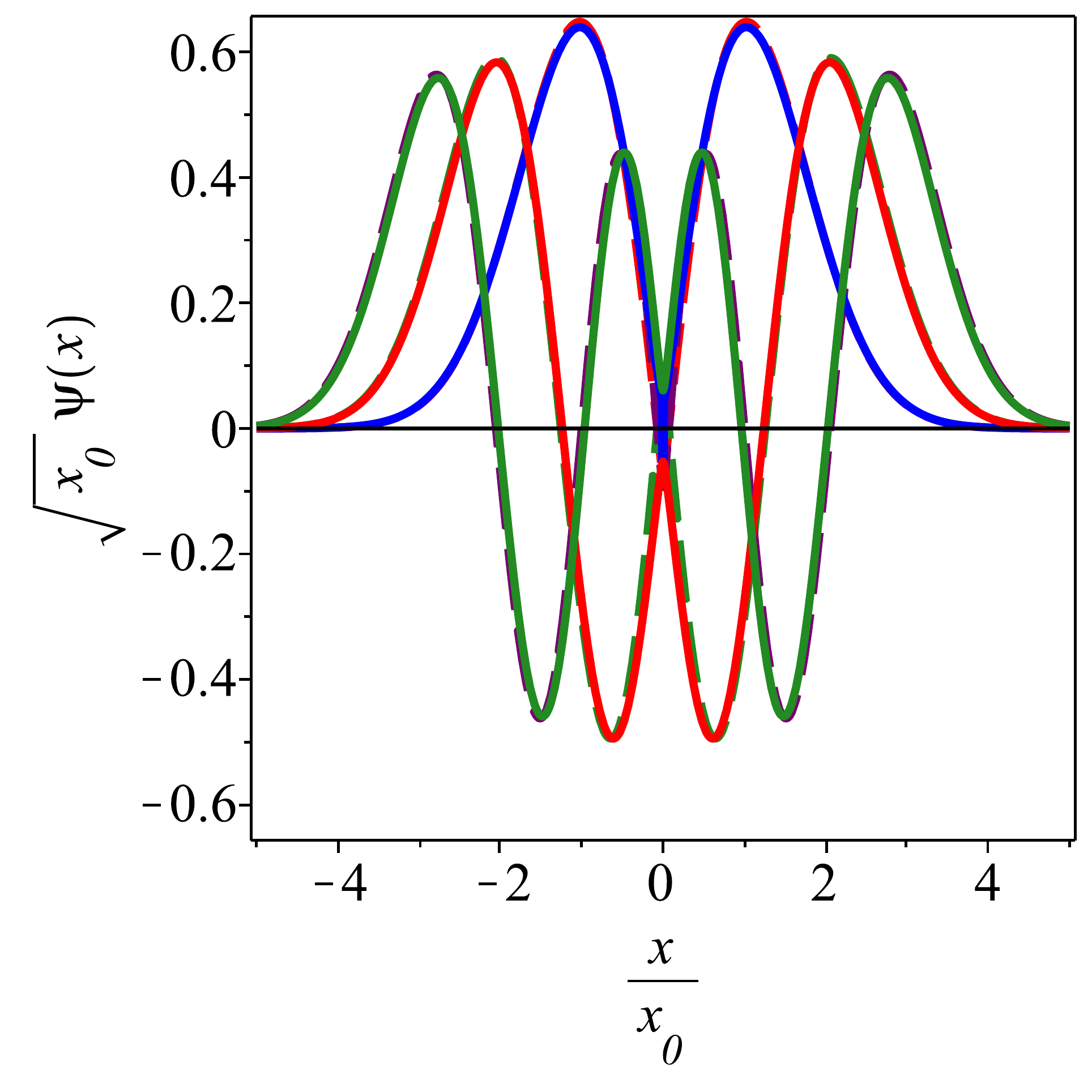}
\caption{The even-parity wave functions for the regularized potential with $\delta=0.01$. The dashed curves correspond to $\alpha=-0.1$ while the solid curves are for $\alpha=0.1$. 
For positive $\alpha$, the colour scheme is $n_{\even}=0$ (blue), $n_{\even}=1$ (red), and $n_{\even}=2$ (green), 
whereas for negative $\alpha$, $n_{\even}=1$ (red), $n_{\even}=2$ (green), and $n_{\even}=3$ (purple).}
\label{fig:RegPsiEven}
\end{figure}

Let us now turn to the wave function for even parity-states, given in Eq.~\eqref{eq:PsiEven}. 
Continuity of $\psi$ at $x=\delta x_{0}$ imposes the condition
\begin{equation}
\label{eq:AIEven}
A_{\text{I}}\cos\left(q\delta x_{0}\right)=B_{\text{II}}\delta^{\nu}e^{-\delta^{2}/2}U\left(\frac{\nu-\kappa}{2},\nu+\frac{1}{2},\delta^{2}\right).
\end{equation}
Normalization of $\psi$ requires $\int_{-\infty}^{\infty}dx\left|\psi\left(x\right)\right|^{2}=1$. 
After inserting Eq.~\eqref{eq:AIEven} into Eq.~\eqref{eq:PsiEven}, then performing the normalization integral and solving for $A^{2}_{\text{I}}$, we obtain
\begin{align}
A^{2}_{\text{I}}&=\frac{1}{2\delta x_{0}}\Biggl\{ \int_{0}^{1}dy\cos^{2}(q\delta x_{0}y)+\cos^{2}(q\delta x_{0})\int_{1}^{\infty}dyy^{2\nu}\nonumber\\
&\quad\times e^{-\delta^{2}(y^{2}-1)}\left[\frac{U\left(\frac{\nu-\kappa}{2},\nu+\frac{1}{2},\delta^{2}y^{2}\right)}{U\left(\frac{\nu-\kappa}{2},\nu+\frac{1}{2},\delta^{2}\right)}\right]^{2}\Biggr\} ^{-1}.
\end{align}
The first few even-parity wave functions are shown in Fig.~\ref{fig:RegPsiEven}. 
For completeness, we present the results for negative and positive $\alpha$. The analysis for the $\alpha>0$ case is presented in Sec.~\ref{sec:RegSE2EvenPosAlpha}.
As in the previous section, the wave functions strongly resemble those obtained for the
unregularized potential shown in Fig.~\ref{fig:UnregPsiEven}. The agreement improves with smaller values of $\delta$ (not shown)
but the convergence towards the unregularized result is slower than for the odd-parity wave functions.
We now have a complete description of the energy eigenvalues and the wave functions for the even and odd-parity solutions with positive energy.

\section{Ground-state solution with infinite negative energy}
\label{sec:GS}

\subsection{Energy eigenvalue}
\label{sec:GSEnergy}

For the even-parity solutions, Fig.~\ref{fig:RegEigs} shows that there is a state whose energy becomes increasingly negative as $\delta\rightarrow0$. Let us now determine the analytical properties of this solution. 
The eigenvalue condition for even-parity states is given in Eq.~\eqref{eq:EVCEven}. 
Using the identity in Eq.~\eqref{eq:ASIdentity2}, this condition can be expressed as 
\begin{equation}
\label{eq:EVCEven2}
q\delta x_{0}\tan\left(q\delta x_{0}\right) = -\delta^{2}-\nu+2\delta^{2}\frac{U\left(\frac{\nu-\kappa}{2},\nu+\frac{3}{2},\delta^{2}\right)}{U\left(\frac{\nu-\kappa}{2},\nu+\frac{1}{2},\delta^{2}\right)}.
\end{equation}
From Eq.~(13.8.11) of Ref.~\cite{NIST2020}, we have
\begin{align}
\label{eq:UAsymptote}
& \lim_{a\rightarrow\infty}U\left(a,b,z\right)  = 2\left(\frac{z}{a}\right)^{\frac{1}{2}\left(1-b\right)}\frac{e^{z/2}}{\Gamma\left(a\right)}\biggl\{K_{b-1}\left(2\sqrt{az}\right)\nonumber\\
& \times\sum_{s=0}^{\infty}\frac{p_{s}\left(b,z\right)}{a^{s}}+\sqrt{\frac{z}{a}}K_{b}\left(2\sqrt{az}\right)\sum_{s=0}^{\infty}\frac{q_{s}\left(b,z\right)}{a^{s}}\biggr\}. 
\end{align}
The $p$ and $q$ coefficients are defined in Eqs.~(13.8.15-13.8.16) of Ref.~\cite{NIST2020}. Here we define $a=\frac{1}{2}\left(\nu-\kappa\right),b=\nu+\frac{1}{2}$, and $z=\delta^{2}$.
Since we are interested in the limit $\delta\rightarrow0$, we need to consider only the $z=0$ values of the first few $p$ and $q$ coefficients, which are given by
\begin{align}
\label{eq:P0Coeff} p_{0}\left(b,z\right) & = 1,\\
p_{1}\left(b,0\right) & = -\frac{b}{2}\left(b-1\right),\\
\label{eq:Q0Coeff} q_{0}\left(b,0\right) & = \frac{b}{2}.
\end{align}
Numerical results indicate that, as $\delta\rightarrow0$, the quantity $\kappa\delta^2$ is constant. This motivates the following series expansion for $\kappa$, as a function of powers of $\delta^{2}$:
\begin{equation}
\label{eq:KAnsatz}
\kappa=-\frac{2c_{0}}{\delta^{2}}+c_{1}-\frac{1}{2}+c_{2}\delta^2+\dots.
\end{equation}

By inserting this ansatz for $\kappa$ in Eq.~\eqref{eq:EVCEven2}, and solving order by order in powers of $\delta^2$, the coefficients $c_{0}$, $c_{1}$, etc., can be deduced. 
The most important coefficients are $c_{0}$ and $c_{1}$, because they appear in expressions that do not vanish as $\delta\rightarrow0$. 
The derivation is lengthy, thus we defer the technical details to Appendix~\ref{App:CCoeffs} and here we present just the final results. 
The coefficient $c_{0}$ is the solution of the following transcendental equation:
\begin{align}
\label{eq:C0Eqn}
c_{0}&=\frac{1}{4}\biggl\{ \left[\sqrt{\left|\alpha\right|-4c_{0}}\tan\left(\sqrt{\left|\alpha\right|-4c_{0}}\right)+\nu\right]\nonumber\\
&\quad\times\frac{K_{\nu-\frac{1}{2}}\left(2\sqrt{c_{0}}\right)}{K_{\nu+\frac{1}{2}}\left(2\sqrt{c_{0}}\right)}\biggr\} ^{2}.
\end{align}
The coefficient $c_{1}$ is determined in closed form to be:
\begin{equation}
\label{eq:C1Eqn}
c_{1}=0.
\end{equation}
The energy is $E=\left(\kappa+\frac{1}{2}\right)\hbar\omega$. Thus, using the previous results, the expansion of the ground-state energy $E_{0}$ in powers of $\delta^{2}$ is
\begin{equation}
\label{eq:GSE}
\frac{E_{0}}{\hbar\omega}=-\frac{2c_{0}}{\delta^{2}}+O\left(\delta^{2}\right).
\end{equation}
Interestingly, notice that there is no constant term in the energy as $\delta\rightarrow0$. 

\begin{figure}[t]
\centering
\includegraphics[scale=0.4]{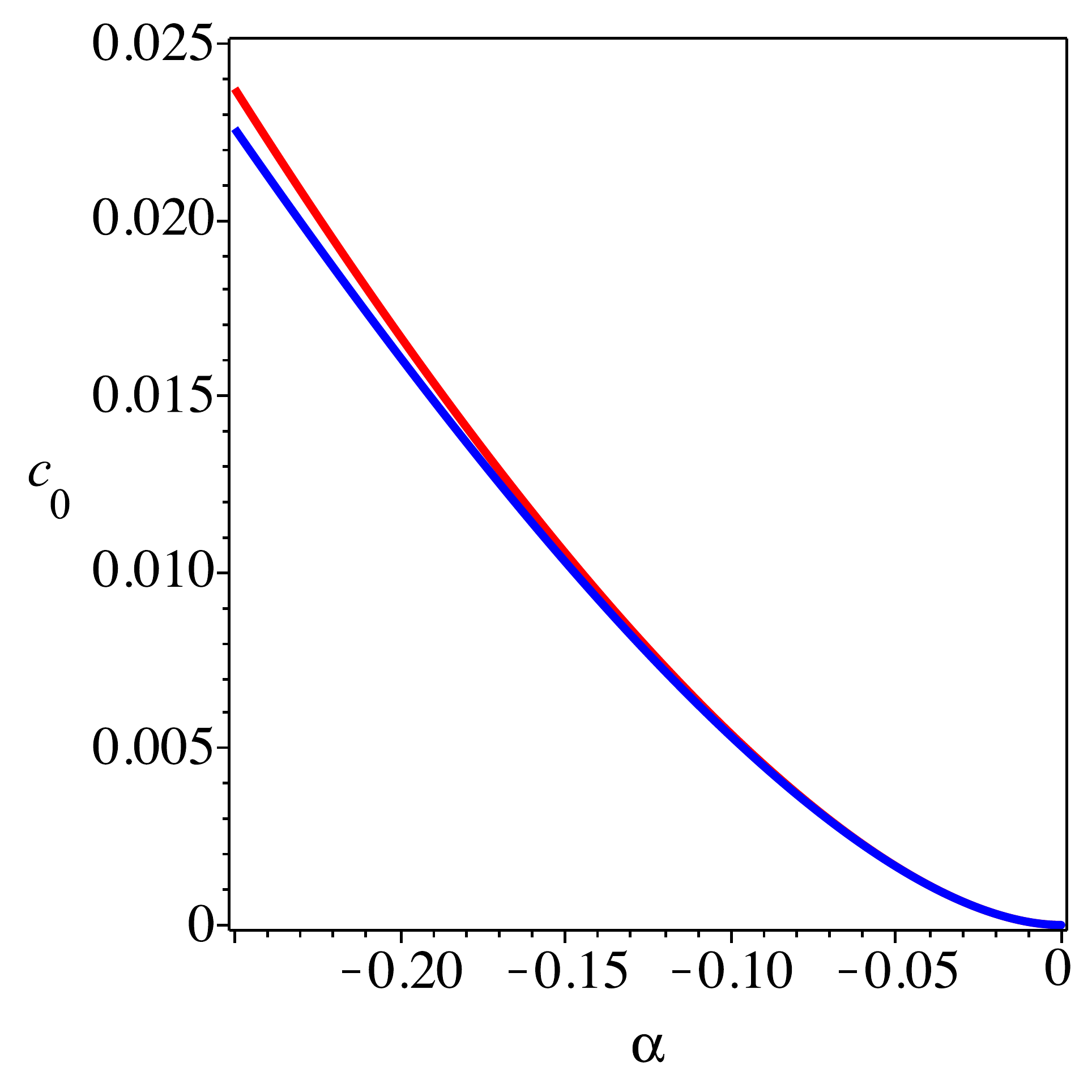}
\caption{The coefficient $c_{0}$ as a function of $\alpha$ for $\alpha\in[-0.25,0)$. The blue curve is obtained by solving the self-consistent equation in Eq.~\eqref{eq:C0Eqn} and the red curve is the approximate analytical result in Eq.~\eqref{eq:C0Eqn2}.}
\label{fig:C0}
\end{figure}

In principle, Eq.~\eqref{eq:C0Eqn} can be numerically solved to determine $c_{0}$ for all $-1/4\leq\alpha<0$ and arbitrary $\delta\ll1$. 
Once $c_{0}$ is deduced, the ground-state energy is then determined from Eq.~\eqref{eq:GSE}. 
Nevertheless, it is preferable to determine a closed-form expression for $c_{0}$ as a function of $\alpha$, applicable in the limit $\delta\ll1$. 
In Appendix~\ref{App:C0Sol} we perform such an analysis. The final result is
\begin{align}
\label{eq:C0Eqn2}
c_{0} & \approx  \Biggl\{\left[1-\frac{2\sqrt{\frac{1}{4}+\alpha}}{\sqrt{\left|\alpha\right|}\tan\sqrt{\left|\alpha\right|}+\frac{1}{2}+\sqrt{\frac{1}{4}+\alpha}}\right] \nonumber\\
&\quad\times \frac{\Gamma\left(1+\sqrt{\frac{1}{4}+\alpha}\right)}{\Gamma\left(1-\sqrt{\frac{1}{4}+\alpha}\right)}
\Biggr\}^{\frac{1}{\sqrt{\frac{1}{4}+\alpha}}}.
\end{align}
This expression is valid provided $4c_{0}\ll1+\sqrt{\frac{1}{4}+\alpha}$.
In Fig.~\ref{fig:C0}, we plot $c_{0}$ as a function of $\alpha$ using both the self-consistent equation in Eq.~\eqref{eq:C0Eqn} (blue curve)
and the approximate formula in Eq.~\eqref{eq:C0Eqn2} (red curve).
As can be observed in this figure, the analytical result gives a very good approximation for nearly the entire range of values of $\alpha$. 
It is only in the limiting case $\alpha\rightarrow-1/4$ that the approximate result deviates noticeably from the exact result.
As $\alpha\rightarrow-1/4$, Eq.~\eqref{eq:C0Eqn} gives $4c_{0}\approx0.0904$, whereas Eq.~\eqref{eq:C0Eqn2} gives $4c_{0}\approx0.0949$; 
these values are not extremely small compared to unity, which is what is required for the assumption $4c_{0}\ll1+\sqrt{\frac{1}{4}+\alpha}$ to be valid. 
For very small and negative $\alpha$ ($\alpha<0$ and $\left|\alpha\right|\ll1$), Eq.~\eqref{eq:C0Eqn2} reduces to 
\begin{equation}
c_{0}\approx\left|\alpha\right|^{\frac{2}{\sqrt{1 - 4|\alpha|}}}.
\end{equation}

\begin{figure}[t]
\centering
\includegraphics[scale=0.4]{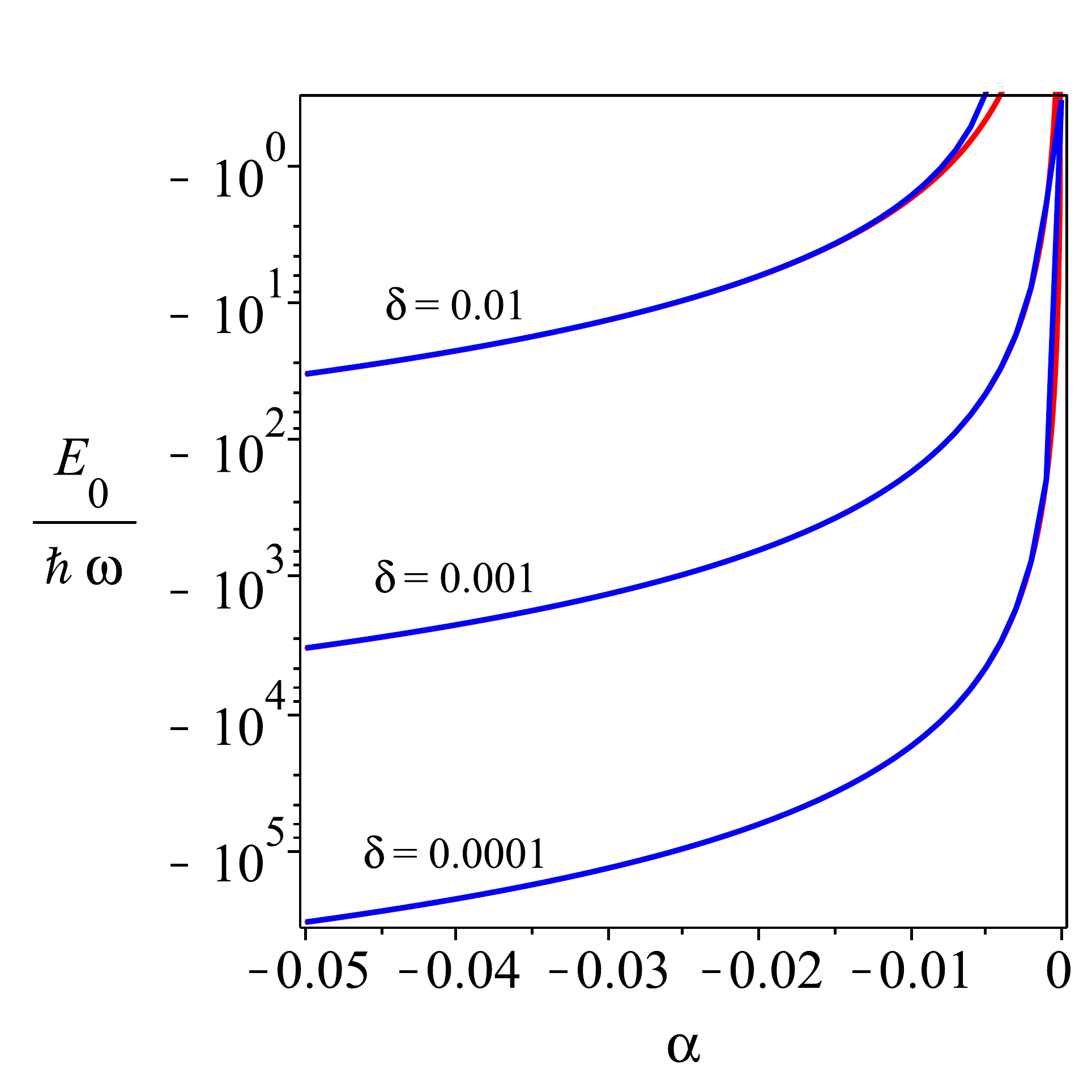}
\caption{The (even-parity) ground-state energy as a function of $\alpha$ for negative values of $\alpha$. The blue curves correspond to the exact energy determined by solving Eq.~\eqref{eq:EVCEven}, whereas the red curves are obtained using Eqs.~\eqref{eq:GSE} and \eqref{eq:C0Eqn2}. 
In order from top to bottom, the figures have $\delta=0.01,0.001,$ and $0.0001$ respectively.}
\label{fig:GSEigs}
\end{figure}

In Fig.~\ref{fig:GSEigs}, we provide a plot of the ground-state energy on a logarithmic scale as a function of $\alpha$ for a small,
negative range near zero, and for three different values of the regularization parameter $\delta$. 
The ground-state energy decreases significantly with increasing $|\alpha|$ and with decreasing $\delta$. 
The approximate result from Eqs.~\eqref{eq:GSE} and \eqref{eq:C0Eqn2} is also shown (in red), and it is indistinguishable from the exact result for most of the range shown.

\subsection{Ground-state wave function}
\label{sec:GSPsi}

The wave function for even-parity solutions is given in Eq.~\eqref{eq:PsiEven}. 
To determine the form of $\psi$ in the limit $\kappa \rightarrow -\infty$ ($\left|\kappa\right|\rightarrow\infty$), we use the identity in Eq.~\eqref{eq:UAsymptote}, keeping only the term with $p_{0}$ as its coefficient. 
Applying this identity to Eq.~\eqref{eq:PsiEven}, the wave function becomes 
\begin{equation}
\psi\left(x\right) = N\lim_{\left|\kappa\right|\rightarrow\infty}\left|\frac{x}{x_{0}}\right|^{\frac{1}{2}}K_{\nu-\frac{1}{2}}\left(\sqrt{2\left|\kappa\right|}\left|\frac{x}{x_{0}}\right|\right).
\end{equation}
Here, $N$ denotes the normalization constant. 
The asymptotic behaviour of the order $\lambda$ modified Bessel function of the second kind is (see Eq.~(9.7.2) of Ref.~\cite{AbramowitzStegun}):
\begin{equation}
K_{\lambda}\left(z\right)\sim\sqrt{\frac{\pi}{2z}}e^{-z},\quad z\rightarrow\infty.
\end{equation}
Thus, we have 
\begin{equation}
\psi\left(x\right)=\lim_{\left|\kappa\right|\rightarrow\infty}Ne^{-\sqrt{2\left|\kappa\right|}\left|x\right|/x_{0}}.
\end{equation}
The normalization constant is determined as usual:
\begin{equation}
1 = \int_{-\infty}^{\infty}dx\left|\psi\left(x\right)\right|^{2} =  \frac{N^{2}x_{0}}{\sqrt{2\left|\kappa\right|}}.
\end{equation}

\begin{figure}[t]
\centering
\includegraphics[scale=0.4]{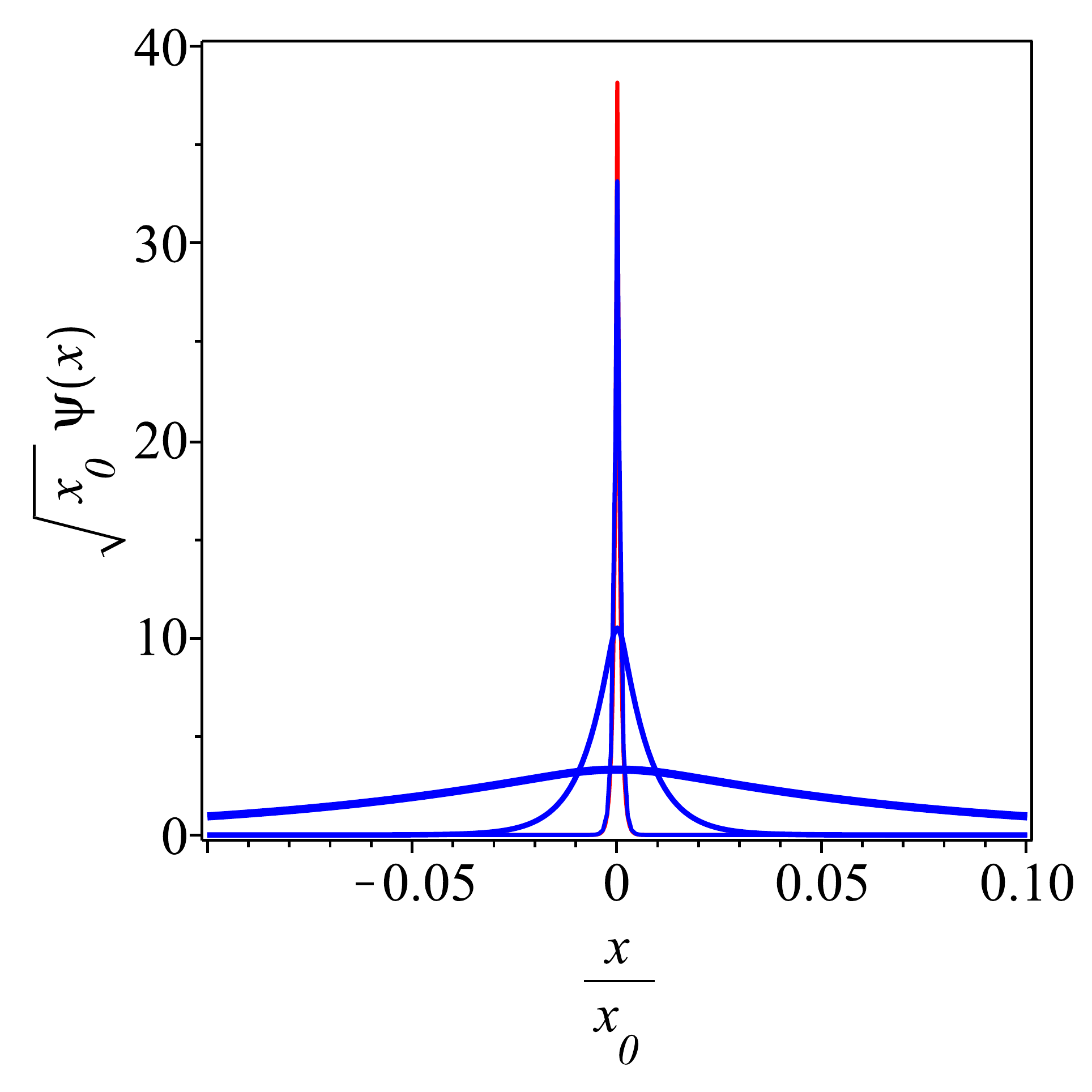}
\caption{The ground-state wave function for $\alpha=-0.1$. The blue curves are obtained using the exact result in Eq.~\eqref{eq:PsiEven}. The thick line (broadest curve) has $\delta=0.01$, the medium line has $\delta=0.001$, and the thin line (narrowest curve)
has $\delta=0.0001$. The red curve is obtained using Eq.~\eqref{eq:GSWavefunction}, where we used the value of $\kappa$ obtained from Eq.~\eqref{eq:EVCEven} with $\delta=0.0001$. As $\delta$ becomes smaller the curve becomes increasingly sharp and the probability density limits to a sharply spiked function, as indicated in Eq.~\eqref{eq:GSProbDensity}.}  
\label{fig:PsiGS}
\end{figure}

Thus, the normalized ground-state wave function (in the limit $\delta\ll1$) is
\begin{equation}
\label{eq:GSWavefunction}
\psi\left(x\right)=\lim_{\left|\kappa\right|\rightarrow\infty}\left(\frac{2\left|\kappa\right|}{x_{0}^{2}}\right)^{\frac{1}{4}}e^{-\sqrt{2\left|\kappa\right|}\left|x\right|/x_{0}}.
\end{equation}
Interestingly, this ground-state wave function has the same functional form as that of the ground-state wave function for the 1D hydrogen atom~\cite{Boyack2021}. 
Indeed, if we replace the length scale $x_{0}$ by the Bohr radius $a_{0}$, and replace $\sqrt{2|\kappa|}$ by the parameter $1/\beta$, where the condition $|\kappa|\rightarrow\infty$ now becomes $\beta\rightarrow0$, then we recover the ground-state wave function for the 1D hydrogen atom~\cite{Boyack2021}.
Notice that the probability density limits to a Dirac delta function:
\begin{equation}
\label{eq:GSProbDensity}
\left|\psi\left(x\right)\right|^{2}=\lim_{\left|\kappa\right|\rightarrow\infty}\sqrt{\frac{2\left|\kappa\right|}{x_{0}^{2}}}e^{-2\sqrt{2\left|\kappa\right|}\left|x\right|/x_{0}}=\delta\left(x\right).
\end{equation}
Here we used the definition 
\begin{equation}
\delta\left(x\right)=\lim_{\epsilon\rightarrow0}\frac{1}{\sqrt{\epsilon}}e^{-2\left|x\right|/\sqrt{\epsilon}}.
\end{equation}
In Fig.~\ref{fig:PsiGS}, we plot the exact ground-state wave function obtained using Eq.~\eqref{eq:PsiEven} for various values of $\delta$, and we also plot the limiting wave function Eq.~\eqref{eq:GSWavefunction}. 
There is good agreement between the exact result (blue) and the approximate wave function (red) for $\delta$ very small.

\section{Regularized potential case ii: $\alpha>0$}
\label{sec:RegSE2}

For $\alpha>0$, the minimum of the potential is now $V_{\text{min}}>0$.
Thus, the range of permissible energies is $0<E<\infty$. 
As a result, the state that has infinite negative energy in the case $-1/4\leq\alpha<0$ will now have a finite and positive energy for $\alpha>0$. 
We now investigate the odd and even-parity eigenfunctions as in the previous sections. 

\subsection{Odd-parity solutions}
\label{sec:RegSE2OddPosAlpha}

Since $\alpha>0$, the potential $\widetilde{V}_{0}\rightarrow+\infty$ as $\delta\rightarrow0$.
Thus, we define $k^{2}=\frac{2m}{\hbar^{2}}\left(\widetilde{V}_{0}-E\right)$, where E is defined in Eq.~\eqref{eq:EandyDefs}. 
Using the definitions of $k$ and $E$, the quantity $k\delta x_{0}$ can be expressed in terms of $\kappa$ as follows:
\begin{equation}
\left(k\delta x_{0}\right)^{2}=\left(\delta^{4}+\alpha\right)-\left(2\kappa+1\right)\delta^{2}
\end{equation}
In region I, we have
\begin{equation}
\label{eq:SERegionI_Eq_PosAlpha}
\psi^{\prime\prime}\left(x\right)-k^{2}\psi\left(x\right)=0.
\end{equation}
The solutions are 
\begin{equation}
\label{eq:SERegionI_Sol_PosAlpha}
\psi_{\text{I}}\left(x\right)=A_{\text{I}}\cosh\left(kx\right)+B_{\text{I}}\sinh\left(kx\right).
\end{equation}
Let us first consider the odd-parity solutions: $A_{\text{I}}=0$. 
In region II the Schr$\ddot{\text{o}}$dinger equation is the same as in the previous section. 
The wave function is then 
\begin{equation}
\psi\left(x\right)=\left\{ \begin{array}{c}
\qquad \qquad B_{\text{I}}\sinh\left(kx\right),\hspace{14mm} x\leq\delta x_{0}\\
B_{\text{II}}y^{\nu}e^{-\frac{1}{2}y^{2}}U\left(\frac{\nu-\kappa}{2},\nu+\frac{1}{2},y^{2}\right),\ x\geq\delta x_{0}.
\end{array}\right.\label{eq:PsiOdd_PosAlpha}
\end{equation}
For $\alpha>0$, $\nu$ is again taken as in Eq.~\eqref{eq:Nu}, which means that $\nu>1$. We again define $\kappa$ as in Eq.~\eqref{eq:KEqn}. 
The energy eigenvalues are determined from the continuity of $\psi$ at $x=\delta x_{0}$, which gives
\begin{equation}
\label{eq:EVCOdd_PosAlpha}
k\delta x_{0}\coth\left(k\delta x_{0}\right)=\delta^{2}-\kappa-1-2\frac{U\left(\frac{\nu-\kappa}{2}-1,\nu+\frac{1}{2},\delta^{2}\right)}{U\left(\frac{\nu-\kappa}{2},\nu+\frac{1}{2},\delta^{2}\right)}.
\end{equation}

As $\delta\rightarrow0$, $\left(k\delta x_{0}\right)^{2}\rightarrow\alpha$.
Thus, for the odd-parity solutions, we now have 
$k\delta x_{0}\coth\left(k\delta x_{0}\right)\rightarrow\sqrt{\alpha}\coth\sqrt{\alpha}$.
The correction term $\epsilon_{n}$ can be determined by following the analogous derivation given in Appendix~\ref{App:OddError} for the case $-1/4\leq\alpha<0$. 
The only difference is the replacement of the $\cot$ function by the $\coth$ function. Thus, the final result is
\begin{align}
\label{eq:OddError2}
\epsilon_{n}&=\left(\frac{\nu-\sqrt{\alpha}\coth\sqrt{\alpha}}{\nu-1+\sqrt{\alpha}\coth\sqrt{\alpha}}\right)\nonumber\\
&\quad\times \frac{\left(-1\right)^{n}}{\Gamma\left(\frac{1}{2}-\nu-n\right)n!}\frac{\Gamma\left(\frac{3}{2}-\nu\right)}{\Gamma\left(\nu+\frac{1}{2}\right)}\delta^{2\nu-1}.
\end{align}
Since $\nu>1/2$, $\epsilon_{n}\rightarrow0$ as $\delta\rightarrow0$. 

Let us now turn to the wave function for odd-parity states, given in Eq.~\eqref{eq:PsiOdd_PosAlpha}.
Continuity of $\psi$ at $x=\delta x_{0}$ imposes the condition
\begin{equation}
\label{eq:BIIOdd_PosAlpha}
B_{\text{I}}\sinh\left(k\delta x_{0}\right)=B_{\text{II}}\delta^{\nu}e^{-\delta^{2}/2}U\left(\frac{\nu-\kappa}{2},\nu+\frac{1}{2},\delta^{2}\right).
\end{equation}
Normalization of $\psi$ requires $\int_{-\infty}^{\infty}dx\left|\psi\left(x\right)\right|^{2}=1$. 
After inserting Eq.~\eqref{eq:BIIOdd_PosAlpha} into Eq.~\eqref{eq:PsiOdd_PosAlpha}, then performing the normalization integral and solving for $B^{2}_{\text{I}}$, we obtain
\begin{align}
B^{2}_{\text{I}}&=\frac{1}{2\delta x_{0}}\Biggl\{ \int_{0}^{1}dy\sinh^{2}\left(k\delta x_{0}y\right)+\sinh^{2}\left(k\delta x_{0}\right)\int_{1}^{\infty}dyy^{2\nu}\nonumber\\
&\quad\times e^{-\delta^{2}\left(y^{2}-1\right)}\left[\frac{U\left(\frac{\nu-\kappa}{2},\nu+\frac{1}{2},\delta^{2}y^{2}\right)}{U\left(\frac{\nu-\kappa}{2},\nu+\frac{1}{2},\delta^{2}\right)}\right]^{2}\Biggr\}^{-1}.
\end{align}
The first few odd-parity wave functions are shown in Fig.~\ref{fig:RegPsiOdd}. In the next section we investigate the even-parity solutions.

\subsection{Even-parity solutions}
\label{sec:RegSE2EvenPosAlpha}

\begin{figure}[t]
\centering
\includegraphics[scale=0.4]{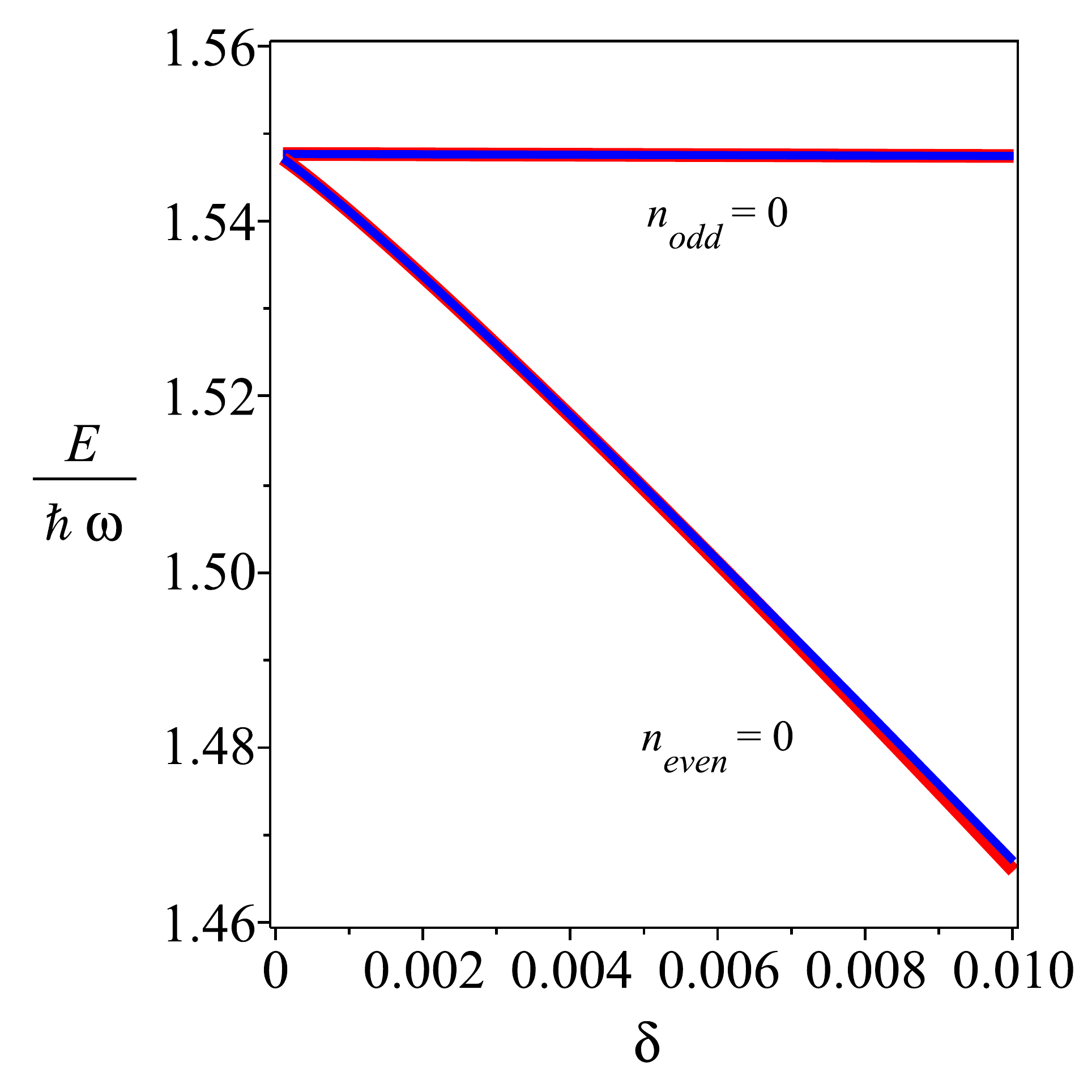}
\caption{Exact energy eigenvalues for the even and odd solutions (blue) versus the approximate energy eigenvalues (red). 
We have made the red curve thicker so that it is visible, since the energies are extremely close, particularly in the case of the odd solutions. Here $\alpha=0.05$.}
\label{fig:EnergyCorrPosAlpha}
\end{figure}

As mentioned at the start of this section, there is no infinite negative energy state for $\alpha>0$. 
The Schr$\ddot{\text{o}}$dinger equation in region I is given in Eq.~\eqref{eq:SERegionI_Eq_PosAlpha}, with the general solution given in Eq.~\eqref{eq:SERegionI_Sol_PosAlpha}. 
For even-parity solutions we set $B_{\text{I}}=0$. The wave function is then
\begin{equation}
\label{eq:PsiEven_PosAlpha}
\psi\left(x\right)=\left\{ \begin{array}{c}
\qquad \qquad A_{\text{I}}\cosh\left(kx\right),\hspace{13mm} x\leq\delta x_{0}\\
B_{\text{II}}y^{\nu}e^{-\frac{1}{2}y^{2}}U\left(\frac{\nu-\kappa}{2},\nu+\frac{1}{2},y^{2}\right),\ x\geq\delta x_{0}.
\end{array}\right.
\end{equation}
The eigenvalue condition is determined by requiring continuity of $\psi^{\prime}/\psi$ at $x=\delta x_{0}$. 
The final result, in contrast to Eq.~\eqref{eq:EVCOdd_PosAlpha} for the odd-parity solutions, is given by
\begin{equation}
k\delta x_{0}\tanh\left(k\delta x_{0}\right)=\delta^{2}-\kappa-1-2\frac{U\left(\frac{\nu-\kappa}{2}-1,\nu+\frac{1}{2},\delta^{2}\right)}{U\left(\frac{\nu-\kappa}{2},\nu+\frac{1}{2},\delta^{2}\right)}.
\label{eq:EVCEven_PosAlpha}
\end{equation}
The correction term $\epsilon_{n}$ can be determined by following the analogous derivation given in Appendix~\ref{App:OddError} for the case $-1/4\leq\alpha<0$. 
The only difference is the replacement of the $\cot$ function by the $\tanh$ function. Thus, the final result is
\begin{align}
\label{eq:EvenError2}
\epsilon_{n}&=\left(\frac{\nu-\sqrt{\alpha}\tanh\sqrt{\alpha}}{\nu-1+\sqrt{\alpha}\tanh\sqrt{\alpha}}\right)\nonumber\\
&\quad\times\frac{\left(-1\right)^{n}}{\Gamma\left(\frac{1}{2}-\nu-n\right)n!}\frac{\Gamma\left(\frac{3}{2}-\nu\right)}{\Gamma\left(\nu+\frac{1}{2}\right)}\delta^{2\nu-1}.
\end{align}
In Fig.~\ref{fig:EnergyCorrPosAlpha}, we compare the exact energy eigenvalues (shown in blue) computed using Eqs.~\eqref{eq:EVCOdd_PosAlpha} and \eqref{eq:EVCEven_PosAlpha} 
for odd and even solutions respectively, against those determined using Eqs.~\eqref{eq:KEqn}, \eqref{eq:OddError2}, and \eqref{eq:EvenError2} (shown in red) for small values of $\delta$ and $\alpha=0.05$. 
The results are in very good agreement, as they were for $\alpha<0$. 

Let us now turn to the wave function for even-parity states, given in Eq.~\eqref{eq:PsiEven_PosAlpha}.
Continuity of $\psi$ at $x=\delta x_{0}$ imposes the condition
\begin{equation}
\label{eq:AIEven_PosAlpha}
A_{\text{I}}\cosh\left(k\delta x_{0}\right)=B_{\text{II}}\delta^{\nu}e^{-\delta^{2}/2}U\left(\frac{\nu-\kappa}{2},\nu+\frac{1}{2},\delta^{2}\right).
\end{equation}
Normalization of $\psi$ requires $\int_{-\infty}^{\infty}dx\left|\psi\left(x\right)\right|^{2}=1$. 
After inserting Eq.~\eqref{eq:AIEven_PosAlpha} into Eq.~\eqref{eq:PsiEven_PosAlpha}, then performing the normalization integral and solving for $A^{2}_{\text{I}}$, we obtain
\begin{align}
A^{2}_{\text{I}}&=\frac{1}{2\delta x_{0}}\Biggl\{ \int_{0}^{1}dy\cosh^{2}(q\delta x_{0}y)+\cosh^{2}(q\delta x_{0})\int_{1}^{\infty}dyy^{2\nu}\nonumber\\
&\quad\times e^{-\delta^{2}\left(y^{2}-1\right)}\left[\frac{U\left(\frac{\nu-\kappa}{2},\nu+\frac{1}{2},\left(\delta y\right)^{2}\right)}{U\left(\frac{\nu-\kappa}{2},\nu+\frac{1}{2},\delta^{2}\right)}\right]^{2}\Biggr\} ^{-1}.
\end{align}
The wave functions are shown in Fig.~\ref{fig:RegPsiEven}. 
Note that these results again look very much like their counterparts with $\alpha = -0.1$ (also shown in Fig.~\ref{fig:RegPsiEven}).
Indeed, both odd and even-parity wave functions will converge towards the unregularized solutions shown in 
Figs.~\ref{fig:UnregPsiOdd} and \ref{fig:UnregPsiEven}, as $\delta$ is taken smaller and smaller. 
The sign of $\alpha$ becomes immaterial. 
For $\alpha > 0$ these states are excluded from the barrier region by the barrier itself. For $\alpha < 0$ the same
set of states are excluded from this same region by the pseudopotential barrier~\cite{Ibrahim2018} created by the ground-state wave function. 
The remarkable result here is that even the unregularized potential with a negative value of $\alpha$, for which {\it no negative-energy ground state exists}, has the same behaviour. The higher energy solutions, in the unregularized case, appear to know of the presence
of a state with negative (and infinite!) energy.

\section{Matrix mechanics method for the regularized potential}
\label{sec:matrix_mechanics}

As an additional check of our analytical work on the regularized potential, it is possible to formulate a solution in terms of matrix mechanics~\cite{Marsiglio2009,Nguyen2020}. 
We proceed by embedding the potential given by Eq.~(\ref{eq:Potential1})
in an infinite square well potential (ISW) of width $a$, with $V_{\rm ISW} = 0$ for $0< x < a$ and infinite otherwise. 
This domain is chosen so that we can use a convenient basis set given by
\begin{equation}
\phi_{n}(x) = \sqrt{\frac{2}{a}}\sin\left(\frac{n\pi x}{a}\right), \quad n = 1,2,3,\dots .
\label{basis}
\end{equation}
To make the potential symmetric, we need to shift the potential as well, so that Eq.~(\ref{eq:Potential1}), when regularized, becomes
\begin{equation}
V\left(x\right)=\left\{ \begin{array}{c}
\frac{1}{2}m\omega^{2}\left(x-{a \over 2}\right)^{2}+\frac{\hbar^{2}}{2m}\frac{\alpha}{\left(x-{a\over 2}\right)^{2}},\quad \left|x - {a \over 2}\right| > \epsilon {a \over 2},\\
\frac{1}{2}m\omega^{2}\left(\epsilon{a \over 2}\right)^{2}+\frac{\hbar^{2}}{2m}\frac{\alpha}{(\epsilon{a\over 2})^{2}}, \quad |x - {a \over 2}| < \epsilon {a \over 2}.
\end{array}\right.
\label{eq:regularized_potential_shifted}
\end{equation}
The dimensionless constant $\epsilon$ provides the cutoff; below this cutoff, the potential is replaced by a constant, 
$V_\epsilon \equiv \frac{1}{2}m\omega^{2}(\epsilon{a \over 2})^{2}+\frac{\hbar^{2}}{2m}\frac{\alpha}{\left(\epsilon{a\over 2}\right)^{2}}$, 
as given in the second line in Eq.~(\ref{eq:regularized_potential_shifted}). 
This dimensionless cutoff is related to the cutoff $\delta$, first introduced in Sec.~\ref{sec:RegSE1}, by
\begin{equation}
\epsilon = {2 \over \pi}\sqrt{2 \over \rho}\delta.
\label{eq:fact1}
\end{equation}
The two length scales and the two energy scales are related by
\begin{equation}
{a \over x_0}= \pi \sqrt{\rho \over 2} \ \ \ {\rm and} \ \ \rho \equiv {\hbar \omega \over E_1^{(0)}} \ \ \  {\rm with} \ \ E_1^{(0)} = {\hbar^2 \pi^2 \over 2 m a^2}.
\label{eq:fact2}
\end{equation}
Using the wave function expansion
\begin{equation} 
\psi(x) = \sum_{m=1}^{N_{\rm max}} c_m \phi_m(x),
\label{eq:basis_expansion}
\end{equation}
the usual matrix formulation \cite{Marsiglio2009} results in the matrix equation for the unknown eigenvalue $E$ and eigenvector coefficients $c_n$:
\begin{equation}
\sum_{m=1}^{N_{\rm max}} H_{nm} c_{m} = E c_{n}.
\label{eq:mat_eq}
\end{equation}
Note that care is required to have $N_{\rm max}$ sufficiently large to ensure that errors from the truncated expansion are completely negligible,
and that the infinite square well width $a$ is large enough to ensure none of our results are affected by its presence. 
In practice, the results need to be compiled as a function of both of these parameters, $a$ and $N_{\rm max}$, until convergence is achieved.

The Hamiltonian matrix is divided into three pieces, $H_{nm} = H^{\rm kin}_{nm} + V^{\rm ext}_{nm} + V^{\rm con}_{nm}$:  the kinetic term, the $x$-dependent potential [first line of Eq.~(\ref{eq:regularized_potential_shifted})], 
and the constant potential [second line of Eq.~(\ref{eq:regularized_potential_shifted})], respectively.

Note that, since the potential is even, only matrix elements where $n \pm m$ is even are non zero. We define $p \equiv \pi (n \pm m)$ and ${\rm sinc}(x) \equiv \sin{(x)}/x$. 
We quote the results in units of $E_1^{(0)}$ and use $v_\epsilon \equiv V_\epsilon/E_1^{(0)}$:
\begin{align}
\frac{H^{\text{kin}}_{nm}}{E_{1}^{(0)}} &= \delta_{nm} n^2, \\
\frac{V^{\text{con}}_{nm}}{E_1^{(0)}} &=  \epsilon v_{\epsilon} \Biggl\{\delta_{nm} \left[1 - \left(-1\right)^n \text{sinc}(\pi n \epsilon)\right] \nonumber\\
&\quad + \left(1-\delta_{nm}\right) \left[g_{\epsilon}\left(n-m\right) - g_{\epsilon}\left(n+m\right) \right] \Biggr\}, \\
\frac{V^{\text{ext}}_{nm}}{E_1^{(0)}} &= \left(1 + (-1)^{n+m} \right) \Biggl\{\biggl[\delta_{nm} \left(\frac{\left(1 - \epsilon^3\right)}{24}  - h_{\epsilon}\left(2n\right)\right) \nonumber\\
&\quad + \left(1-\delta_{nm}\right)\left[h_\epsilon(n-m) - h_\epsilon(n+m) \right] \biggr]\frac{\pi^2 \rho^2}{4}  \nonumber \\
&\quad + {\alpha \over \pi^2} \left[k_{\epsilon}\left(n-m\right) - k_{\epsilon}\left(n+m\right)\right] \Biggr\}. 
 \end{align}
The quantities $g_{\epsilon}\left(n\pm m\right), h_{\epsilon}\left(n\pm m\right), k_{\epsilon}\left(n\pm m\right),$ and $\ell_{\epsilon}\left(n\pm m\right)$ are defined by
\begin{align}
g_{\epsilon}\left(n\pm m\right)  & \equiv \cos\left(\frac{p}{2}\right){\rm sinc}\left(\frac{p \epsilon}{2}\right), \\
h_{\epsilon} \left(n\pm m\right) & \equiv \cos\left(\frac{p}{2}\right)\biggl[{2 \over p^3} \sin\left(\frac{p\epsilon}{2}\right) + \frac{1}{p^2}\biggl(\cos{\left(\frac{p}{2}\right)} \nonumber\\
&\quad - \epsilon \cos\left(\frac{p\epsilon }{2}\right) \biggr) - \frac{\epsilon^2}{4p}  \sin\left(\frac{p\epsilon}{2}\right) \biggr], \\
k_{\epsilon} \left(n\pm m\right) & \equiv \cos\left(\frac{p}{2}\right) \left[{2 \over \epsilon}\left(1-\epsilon\right) - \ell_\epsilon \left(n\pm m\right) \right], \\
\ell_{\epsilon} \left(n\pm m\right) & \equiv \int_{\epsilon/2}^{1/2} dx {1 - \cos{\left[\pi \left(n \pm m\right)x\right]} \over x^2} \nonumber \\
 &= {2 \over \epsilon}\left[ 1 - \cos{\left(\frac{p\epsilon}{2}\right)} \right] - 2 \left[ 1 - \cos{\left(\frac{p}{2}\right)} \right] \nonumber \\
&\quad + p\left[ {\rm Si}\left(\frac{p}{2} \right) - {\rm Si}\left(\frac{p \epsilon }{2} \right)  \right]. \label{eq:elleps}
\end{align}
Here, the Sine Integral~\cite{AbramowitzStegun} is defined by
\begin{equation}
{\rm Si}(z) \equiv \int_0^z dt {\sin{t} \over t}.
\label{sine}
\end{equation}

Note that all these quantities are well-defined, but as $\epsilon \rightarrow 0$ [i.e., $\delta \rightarrow 0$ -- see Eq.~(\ref{eq:fact1})] the matrix elements become singular. 
The matrix equation Eq.~(\ref{eq:mat_eq}) can be solved by computer for the eigenvalues and eigenvectors. 
The latter can then be used in Eq.~(\ref{eq:basis_expansion}) to compute the wave functions in real space. 
We can typically use $100 \times 100$ matrices, but as $\delta$ decreases and/or the magnitude of $\alpha$ increases, larger matrices $10000 \times 10000$ or larger are required to achieve convergence. 
Moreover, with both decreasing $\delta$ or increasing $|\alpha|$, Eq.~(\ref{eq:elleps}) becomes more difficult to evaluate accurately. 
We should also emphasize that the length scale $a$ was fabricated for convenience, and the results should not depend on this quantity.
The numerical procedure is also more straightforward for readers not familiar with the properties of the confluent hypergeometric functions.

As a comparison of the numerical approach of this section versus the approach of the previous sections based on hypergeometric functions,
we compute the ground-state energy for fixed $\delta=0.002$ and a range of negative values of $\alpha$. 
The results are shown in Table~\ref{tab:GSEnergy}, for fixed $N_{\rm max} = 10000$ and a length $a$ given by $\rho = 50$ [see
Eq.~(\ref{eq:fact2})]. In practice, this value of $a$ is much larger than required for convergence of just the ground-state energy,
since the ground state is so confined near the origin. For example, for $\alpha = -0.05$ (final row of Table~\ref{tab:GSEnergy}),
with $\rho = 5$ and the same $N_{\rm max} = 10000$, we
obtain $E/(\hbar \omega) = -828.489881$. Comparison with the result attained by the use of Tricomi functions (3rd column)
shows that we can approach the analytical result with arbitrary precision, given sufficient computer power and memory.

\begin{widetext}

\begin{table}[h]
\caption{Comparison between numerical results for the ground-state energy as a function of $\alpha$. Here $\delta=0.002$.
\label{tab:GSEnergy}}
\begin{ruledtabular}
\begin{tabular}{ccccc}
$\alpha$ & Matrix mechanics: \eqref{eq:mat_eq} & Tricomi functions: \eqref{eq:EVCEven} & $-\frac{2c_{0}}{\delta^{2}}$ \eqref{eq:C0Eqn}, \eqref{eq:GSE} & $-\frac{2c_{0}}{\delta^{2}}$ \eqref{eq:GSE}, \eqref{eq:C0Eqn2} \tabularnewline
\hline 
$-0.25$ & $-11294.85744903$ & $-11295.301683$ & $-11295.30170$ & $-11862.24636$ \tabularnewline
$-0.20$ & $-8056.57184873$ & $-8056.826663$ & $-8056.826665$ & $-8353.544755$ \tabularnewline
$-0.15$ & $-5149.73001990$ & $-5149.852852$ & $-5149.852880$ & $-5274.934465$ \tabularnewline
$-0.10$ & $-2679.68183517$ & $-2679.724708$ & $-2679.724730$ & $-2714.801773$ \tabularnewline
$-0.05$ & $-828.48321740$ & $-828.4898894$ & $-828.4900235$ & $-831.9802335$ \tabularnewline
\end{tabular}
\end{ruledtabular}
\end{table}

\end{widetext}

\section{Conclusion}
\label{sec:Conclusion}

In this paper we have performed a thorough analysis of the spectrum of the one-dimensional pseudoharmonic oscillator -- a simple harmonic oscillator in the presence of a $1/x^2$ interaction.
For the case where the potential is unregularized, we have shown that there are doubly-degenerate eigenfunctions when the interaction parameter $\alpha$ is positive, as was already known.
In addition, we have also shown that there are doubly degenerate bound-states in the region $-1/4\leq\alpha<0$. 

We have also studied a regularized version of the pseudoharmonic oscillator, where the interaction is cut off near the origin. 
For this regularized problem, we have again found even and odd-parity eigenfunctions, for $-1/4\leq\alpha<0$ and $\alpha>0$. 
In contrast to the unregularized potential, we have shown that, for $-1/4\leq\alpha<0$, the regularized potential admits a ground-state solution with increasingly negative energy.
We have derived the analytical properties of this ground state and shown that its energy diverges as the inverse square of the cutoff, and that its probability density limits to a Dirac delta function. 
The mathematical features of this solution are analogous to a similar ground state in the regularized one-dimensional hydrogen atom.

The similarity between the regularized and unregularized problems as the regularization parameter $\delta \rightarrow 0$ is, on the one hand, not surprising.  
On the other hand, we do not find a negative energy solution for the unregularized problem with
$-1/4\leq\alpha<0$. Since it is this negative energy ground state that is responsible (through the pseudopotential effect) for 
the properties of the positive-energy solutions, it is in many ways remarkable that the unregularized solutions are so similar to
the regularized solutions. The unregularized problem somehow appears to know about the infinite negative energy ground state. 

One of the remaining areas that requires further investigation is the study of the regularized potential in the regime $\alpha<-1/4$. 
Indeed, bound-state solutions for the regularized potential can be found for $\alpha<-1/4$, since the potential is always finite near the origin. 
It would be desirable to study the behaviour of the ground-state solution with infinite negative energy as $\alpha$ becomes increasingly negative. 
Since it is believed that the unregularized problem does not have finite negative energy bound-state solutions in the 
regime $\alpha<-1/4$, it would be of interest to determine the properties of the eigenfunctions in the regularized problem as the cutoff approaches zero.

\section{Acknowledgments}
We thank J. Lekner for providing insightful comments on this problem. In addition, we also thank E. Dupuis, P. L. S. Lopes, and M. Protter for beneficial discussions.
R.B. was supported by D\'epartement de physique, Universit\'e de Montr\'eal. F.M., A.S., and A.B. were supported in part by the Natural Sciences and Engineering Research Council of Canada (NSERC). 

\appendix
\numberwithin{equation}{section}
\numberwithin{figure}{section}

\section{Derivation of the correction term $\epsilon_{n}$ for odd-parity states where $-1/4\leq\alpha<0$}
\label{App:OddError}

\begin{widetext}
In this appendix we present the derivation of $\epsilon_{n}$ for the odd-parity states in the case where $-1/4\leq\alpha<0$. 
The analysis is similar for $\alpha>0$ and for even-parity states, the only difference being the particular trigonometric or hyperbolic trigonometric functions used.
The Tricomi function $U(a,b,z)$ is defined in Eq.~(13.1.3) of Ref.~\cite{AbramowitzStegun}: 
\begin{equation}
U\left(a,b,z\right)=\frac{\pi}{\sin\left(\pi b\right)}\biggl[\frac{M\left(a,b,z\right)}{\Gamma\left(1+a-b\right)\Gamma\left(b\right)}-z^{1-b}\frac{M\left(1+a-b,2-b,z\right)}{\Gamma\left(a\right)\Gamma\left(2-b\right)}\biggr].
\end{equation}
Here, $M(a,b,z)$ is the Kummer function (also known as the confluent hypergeometric function.) 
Therefore, we have 
\begin{eqnarray}
\frac{U\left(\frac{\nu-\kappa}{2}-1,\nu+\frac{1}{2},\delta^{2}\right)}{U\left(\frac{\nu-\kappa}{2},\nu+\frac{1}{2},\delta^{2}\right)} & = & 
 \frac{\frac{M\left(\frac{\nu-\kappa}{2}-1,\nu+\frac{1}{2},\delta^{2}\right)}{\Gamma\left(\frac{-\nu-\kappa-1}{2}\right)\Gamma\left(\nu+\frac{1}{2}\right)}-\delta^{1-2\nu}\frac{M\left(\frac{-\nu-\kappa-1}{2},\frac{3}{2}-\nu,\delta^{2}\right)}{\Gamma\left(\frac{\nu-\kappa}{2}-1\right)\Gamma\left(\frac{3}{2}-\nu\right)}}{\frac{M\left(\frac{\nu-\kappa}{2},\nu+\frac{1}{2},\delta^{2}\right)}{\Gamma\left(\frac{1-\nu-\kappa}{2}\right)\Gamma\left(\nu+\frac{1}{2}\right)}-\delta^{1-2\nu}\frac{M\left(\frac{1-\nu-\kappa}{2},\frac{3}{2}-\nu,\delta^{2}\right)}{\Gamma\left(\frac{\nu-\kappa}{2}\right)\Gamma\left(\frac{3}{2}-\nu\right)}}.
\end{eqnarray}
We are interested in the limit $\delta\ll1$; thus, we use the series for the hypergeometric function, $M(a,b,z)=1+O(z)$, to approximate this expression as
\begin{equation}
\frac{U\left(\frac{\nu-\kappa}{2}-1,\nu+\frac{1}{2},\delta^{2}\right)}{U\left(\frac{\nu-\kappa}{2},\nu+\frac{1}{2},\delta^{2}\right)}\approx\frac{\frac{1}{\Gamma\left(-\frac{1}{2}-\left(n+\nu+\epsilon_{n}\right)\right)\Gamma\left(\nu+\frac{1}{2}\right)}-\frac{\delta^{1-2\nu}}{\Gamma\left(-\left(n+\epsilon_{n}\right)-1\right)\Gamma\left(\frac{3}{2}-\nu\right)}}{\frac{1}{\Gamma\left(\frac{1}{2}-\left(n+\nu+\epsilon_{n}\right)\right)\Gamma\left(\nu+\frac{1}{2}\right)}-\frac{\delta^{1-2\nu}}{\Gamma\left(-\left(n+\epsilon_{n}\right)\right)\Gamma\left(\frac{3}{2}-\nu\right)}}.
\end{equation}
By using the identities $\Gamma(-x)=-(1+x)\Gamma(-1-x)$ and $\Gamma\left(\frac{1}{2}-x\right)=-\left(\frac{1}{2}+x\right)\Gamma\left(-\frac{1}{2}-x\right)$, we can simplify the expression above to 
\begin{align}
\frac{U\left(\frac{\nu-\kappa}{2}-1,\nu+\frac{1}{2},\delta^{2}\right)}{U\left(\frac{\nu-\kappa}{2},\nu+\frac{1}{2},\delta^{2}\right)} = -\frac{\left(\frac{\kappa+\nu+1}{2}\right)\Gamma\left(-\left(n+\epsilon_{n}\right)\right)\Gamma\left(\frac{3}{2}-\nu\right)-\delta^{1-2\nu}\left(\frac{\kappa-\nu+2}{2}\right)\Gamma\left(\frac{1-\nu-\kappa}{2}\right)\Gamma\left(\nu+\frac{1}{2}\right)}{\Gamma\left(-\left(n+\epsilon_{n}\right)\right)\Gamma\left(\frac{3}{2}-\nu\right)-\delta^{1-2\nu}\Gamma\left(\frac{1-\nu-\kappa}{2}\right)\Gamma\left(\nu+\frac{1}{2}\right)}.
\end{align}
Here we used $n+\nu+\epsilon_{n}=\frac{1}{2}\left(\kappa+\nu\right)$ and $n+\epsilon_{n}=\frac{1}{2}\left(\kappa-\nu\right)$.

Combining this result with Eqs.~\eqref{eq:EVCOdd}-\eqref{eq:KEqn}, the eigenvalue condition is then given by
\begin{align}
\sqrt{\left|\alpha\right|}\cot\sqrt{\left|\alpha\right|}&=\delta^{2}-\kappa-1-2\frac{U\left(-\left(n+\epsilon_{n}\right)-1,\nu+\frac{1}{2},\delta^{2}\right)}{U\left(-\left(n+\epsilon_{n}\right),\nu+\frac{1}{2},\delta^{2}\right)} \nonumber\\
& \approx  -\kappa-1\nonumber \\
 & \quad +2\left\{ \frac{\left(\frac{\kappa+\nu+1}{2}\right)\Gamma\left(-\left(n+\epsilon_{n}\right)\right)\Gamma\left(\frac{3}{2}-\nu\right)-\delta^{1-2\nu}\left(\frac{\kappa-\nu+2}{2}\right)\Gamma\left(\frac{1}{2}-\left(n+\nu+\epsilon_{n}\right)\right)\Gamma\left(\nu+\frac{1}{2}\right)}{\Gamma\left(-\left(n+\epsilon_{n}\right)\right)\Gamma\left(\frac{3}{2}-\nu\right)-\delta^{1-2\nu}\Gamma\left(\frac{1}{2}-\left(n+\nu+\epsilon_{n}\right)\right)\Gamma\left(\nu+\frac{1}{2}\right)}\right\} \nonumber \\
 & = -\nu+\frac{2\nu\Gamma\left(-\left(n+\epsilon_{n}\right)\right)\Gamma\left(\frac{3}{2}-\nu\right)-\delta^{1-2\nu}\Gamma\left(\frac{1}{2}-\left(n+\nu+\epsilon_{n}\right)\right)\Gamma\left(\nu+\frac{1}{2}\right)}{\Gamma\left(-\left(n+\epsilon_{n}\right)\right)\Gamma\left(\frac{3}{2}-\nu\right)-\delta^{1-2\nu}\Gamma\left(\frac{1}{2}-\left(n+\nu+\epsilon_{n}\right)\right)\Gamma\left(\nu+\frac{1}{2}\right)}.
\end{align}
Cross multiply these expressions to obtain
\begin{align}
 &  \left(\sqrt{\left|\alpha\right|}\cot\sqrt{\left|\alpha\right|}+\nu\right)\left[\Gamma\left(-\left(n+\epsilon_{n}\right)\right)\Gamma\left(\frac{3}{2}-\nu\right)-\delta^{1-2\nu}\Gamma\left(\frac{1}{2}-\left(n+\nu+\epsilon_{n}\right)\right)\Gamma\left(\nu+\frac{1}{2}\right)\right]\nonumber\\
 & = 2\nu\Gamma\left(-\left(n+\epsilon_{n}\right)\right)\Gamma\left(\frac{3}{2}-\nu\right)-\delta^{1-2\nu}\Gamma\left(\frac{1}{2}-\left(n+\nu+\epsilon_{n}\right)\right)\Gamma\left(\nu+\frac{1}{2}\right).
\end{align}
Rearrange this equation to get
\begin{equation}
\frac{\Gamma\left(\frac{1}{2}-\left(n+\nu+\epsilon_{n}\right)\right)}{\Gamma\left(-\left(n+\epsilon_{n}\right)\right)}=\left(\frac{\sqrt{\left|\alpha\right|}\cot\sqrt{\left|\alpha\right|}-\nu}{\nu-1+\sqrt{\left|\alpha\right|}\cot\sqrt{\left|\alpha\right|}}\right)\frac{\Gamma\left(\frac{3}{2}-\nu\right)}{\Gamma\left(\nu+\frac{1}{2}\right)}\delta^{2\nu-1}.
\end{equation}
The denominator on the left-hand side of this equation is $\Gamma(-n-\epsilon_{n})=-\pi\left(-1\right)^n/\left[\Gamma\left(n+\epsilon_{n}+1\right)\sin\left(\pi\epsilon_{n}\right)\right]$. In the limit $\epsilon_{n}\ll1$, this becomes $\Gamma(-n-\epsilon_{n})\rightarrow-\left(-1\right)^{n}/\left(n!\epsilon_{n}\right)$. 
In the limit $\delta\ll1$, $\epsilon_{n}\ll1$ and thus we can use the previous result to simplify the equation above to
\begin{equation}
-\Gamma\left(\frac{1}{2}-\nu-n\right)\left(-1\right)^{n}n!\epsilon_{n}=\left(\frac{\sqrt{\left|\alpha\right|}\cot\sqrt{\left|\alpha\right|}-\nu}{\nu-1+\sqrt{\left|\alpha\right|}\cot\sqrt{\left|\alpha\right|}}\right)\frac{\Gamma\left(\frac{3}{2}-\nu\right)}{\Gamma\left(\nu+\frac{1}{2}\right)}\delta^{2\nu-1}.
\end{equation}
Solving this equation for $\epsilon_{n}$ gives the result in Eq.~\eqref{eq:OddError1} of the main text. 

\section{Derivation of the coefficients $c_{0}$ and $c_{1}$}
\label{App:CCoeffs}

The energy eigenvalue equation in Eq.~\eqref{eq:EVCEven2} can be written as 
\begin{equation}
\label{eq:EVCEven3}
q\delta x_{0}\tan\left(q\delta x_{0}\right) = -\delta^{2}-\nu+I,
\end{equation}
where $I$ is the ratio of the two Kummer functions. By using Eq.~\eqref{eq:UAsymptote}, the definitions $a=\frac{1}{2}\left(\nu-\kappa\right), b=\nu+\frac{1}{2}$, and $z=\delta^2$, along with the $p$ and $q$ coefficients in Eqs.~\eqref{eq:P0Coeff}-\eqref{eq:Q0Coeff}, we have
\begin{align}
I & =  2\delta^{2}\frac{U\left(\frac{\nu-\kappa}{2},\nu+\frac{3}{2},\delta^{2}\right)}{U\left(\frac{\nu-\kappa}{2},\nu+\frac{1}{2},\delta^{2}\right)}\nonumber \\
 & =  2\delta^{2}\left(\frac{\delta^{2}}{a}\right)^{-\frac{1}{2}}\frac{\left\{K_{b}\left(2\sqrt{a\delta^{2}}\right)\left[p_{0}\left(b+1,\delta^{2}\right)+\frac{p_{1}\left(b+1,\delta^{2}\right)}{a}\right]+\sqrt{\frac{\delta^{2}}{a}}K_{b+1}\left(2\sqrt{a\delta^{2}}\right)q_{0}\left(b+1,\delta^{2}\right)\right\} }{\left\{ K_{b-1}\left(2\sqrt{a\delta^{2}}\right)\left[p_{0}\left(b,\delta^{2}\right)+\frac{p_{1}\left(b,\delta^{2}\right)}{a}\right]+\sqrt{\frac{\delta^{2}}{a}}K_{b}\left(2\sqrt{a\delta^{2}}\right)q_{0}\left(b,\delta^{2}\right)\right\} }\nonumber \\
 & =  2\delta^{2}\left(\frac{\delta^{2}}{a}\right)^{-\frac{1}{2}}\frac{\left\{ K_{b}\left(2\sqrt{a\delta^{2}}\right)\left[1-\frac{b\left(b+1\right)}{2a}\right]+\sqrt{\frac{\delta^{2}}{a}}K_{b+1}\left(2\sqrt{a\delta^{2}}\right)\frac{b+1}{2}\right\} }{\left\{ K_{b-1}\left(2\sqrt{a\delta^{2}}\right)\left[1-\frac{b\left(b-1\right)}{2a}\right]+\sqrt{\frac{\delta^{2}}{a}}K_{b}\left(2\sqrt{a\delta^{2}}\right)\frac{b}{2}\right\} }.
\end{align}
The ansatz for $\kappa$ is given by 
\begin{equation}
\kappa=-\frac{2c_{0}}{\delta^{2}}+c_{1}-\frac{1}{2}+O\left(\delta^2\right).
\end{equation}
Therefore, 
\begin{align}
\left(\frac{\delta^{2}}{a}\right)^{-\frac{1}{2}} & =  \sqrt{\frac{\nu-\kappa}{2\delta^{2}}}\nonumber \\
 & =  \sqrt{\frac{1}{\delta^{2}}\left(\frac{c_{0}}{\delta^{2}}+\frac{b-c_{1}}{2}\right)}\nonumber \\
 & \approx  \frac{\sqrt{c_{0}}}{\delta^{2}}+\frac{b-c_{1}}{4\sqrt{c_{0}}}.
\end{align}
Similarly, $\sqrt{\frac{\delta^{2}}{a}}\approx\frac{\delta^{2}}{\sqrt{c_{0}}}$.
The argument of the modified Bessel functions is 
\begin{align}
\sqrt{a\delta^{2}} & = \sqrt{\left(\frac{\nu-\kappa}{2}\right)\delta^{2}}\nonumber \\
 & = \sqrt{c_{0}+\frac{1}{2}\left(b-c_{1}\right)\delta^{2}}\nonumber \\
 & \approx \sqrt{c_{0}}+\frac{b-c_{1}}{4\sqrt{c_{0}}}\delta^{2}.
\end{align}
The expansion of the modified Bessel functions, in powers of $\delta^{2}$, is given by
\begin{equation}
K_{\lambda}\left(2\sqrt{a\delta^{2}}\right)=K_{\lambda}\left(2\sqrt{c_{0}}+\frac{b-c_{1}}{2\sqrt{c_{0}}}\delta^{2}\right)=K_{\lambda}\left(2\sqrt{c_{0}}\right)+\left(\frac{b-c_{1}}{2\sqrt{c_{0}}}\right)\delta^{2}K_{\lambda}^{\prime}\left(2\sqrt{c_{0}}\right).
\end{equation}
Here, $\lambda$ is an arbitrary order of the modified Bessel function of the second kind. Finally, note that
\begin{align}
\frac{1}{a} & = \frac{2}{\nu-\kappa}\nonumber \\
 & \approx \frac{\delta^{2}}{c_{0}}.
\end{align}

For convenience, we omit the arguments of the modified Bessel functions, which are all $2\sqrt{c_{0}}$. 
Thus, the quantity $I$ is given by 
\begin{align}
I & =  2\left[\sqrt{c_{0}}+\left(\frac{b-c_{1}}{4\sqrt{c_{0}}}\right)\delta^{2}\right]\frac{\left\{ \left[K_{b}+\left(\frac{b-c_{1}}{2\sqrt{c_{0}}}\right)\delta^{2}K_{b}^{\prime}\right]\left[1-\frac{b\left(b+1\right)}{2}\frac{\delta^{2}}{c_{0}}\right]+\frac{\delta^{2}}{\sqrt{c_{0}}}\left(\frac{b+1}{2}\right)K_{b+1}\right\} }{\left\{ \left[K_{b-1}+\left(\frac{b-c_{1}}{2\sqrt{c_{0}}}\right)\delta^{2}K_{b-1}^{\prime}\right]\left[1-\frac{b\left(b-1\right)}{2}\frac{\delta^{2}}{c_{0}}\right]+\frac{\delta^{2}}{\sqrt{c_{0}}}\frac{b}{2}K_{b}\right\} }\nonumber \\
  & = 2\left[\sqrt{c_{0}}+\left(\frac{b-c_{1}}{4\sqrt{c_{0}}}\right)\delta^{2}\right]\frac{K_{b}}{K_{b-1}}\frac{\left\{ 1+\left(\frac{b-c_{1}}{2\sqrt{c_{0}}}\right)\delta^{2}\frac{K_{b}^{\prime}}{K_{b}}-\frac{b\left(b+1\right)}{2}\frac{\delta^{2}}{c_{0}}+\frac{\delta^{2}}{\sqrt{c_{0}}}\left(\frac{b+1}{2}\right)\frac{K_{b+1}}{K_{b}}\right\} }{\left\{ 1+\left(\frac{b-c_{1}}{2\sqrt{c_{0}}}\right)\delta^{2}\frac{K_{b-1}^{\prime}}{K_{b-1}}-\frac{b\left(b-1\right)}{2}\frac{\delta^{2}}{c_{0}}+\frac{\delta^{2}}{\sqrt{c_{0}}}\frac{b}{2}\frac{K_{b}}{K_{b-1}}\right\} }\nonumber \\
 & =  2\left[\sqrt{c_{0}}+\left(\frac{b-c_{1}}{4\sqrt{c_{0}}}\right)\delta^{2}\right]\frac{K_{b}}{K_{b-1}}\left[1+\left(\frac{b-c_{1}}{2\sqrt{c_{0}}}\right)\delta^{2}\frac{K_{b}^{\prime}}{K_{b}}-\frac{b\left(b+1\right)}{2}\frac{\delta^{2}}{c_{0}}+\frac{\delta^{2}}{\sqrt{c_{0}}}\left(\frac{b+1}{2}\right)\frac{K_{b+1}}{K_{b}}\right]\nonumber \\
 &  \quad\times\left[1-\left(\frac{b-c_{1}}{2\sqrt{c_{0}}}\right)\delta^{2}\frac{K_{b-1}^{\prime}}{K_{b-1}}+\frac{b\left(b-1\right)}{2}\frac{\delta^{2}}{c_{0}}-\frac{\delta^{2}}{\sqrt{c_{0}}}\frac{b}{2}\frac{K_{b}}{K_{b-1}}\right].
\end{align}
Thus, to $O\left(\delta^{2}\right)$, we have 
\begin{align}
I & = 2\sqrt{c_{0}}\frac{K_{b}}{K_{b-1}}+\left(\frac{b-c_{1}}{2}\right)\delta^{2}\frac{K_{b}}{K_{b-1}}\left(\frac{1}{\sqrt{c_{0}}}+2\frac{K_{b}^{\prime}}{K_{b}}-2\frac{K_{b-1}^{\prime}}{K_{b-1}}\right)\nonumber \\
 & \quad+2\delta^{2}\frac{K_{b}}{K_{b-1}}\left[-\frac{b\left(b+1\right)}{2\sqrt{c_{0}}}+\left(\frac{b+1}{2}\right)\frac{K_{b+1}}{K_{b}}+\frac{b\left(b-1\right)}{2\sqrt{c_{0}}}-\frac{b}{2}\frac{K_{b}}{K_{b-1}}\right].
\end{align}
The self-consistent equation in Eq.~\eqref{eq:EVCEven3} now becomes 
\begin{align}
\label{eq:EVCEven4}
q\delta x_{0}\tan\left(q\delta x_{0}\right)+\nu & = -\delta^{2}+I\nonumber \\
 & =  2\sqrt{c_{0}}\frac{K_{b}}{K_{b-1}} -\delta^{2}+\left(\frac{b-c_{1}}{2}\right)\delta^{2}\frac{K_{b}}{K_{b-1}}\left(\frac{1}{\sqrt{c_{0}}}+2\frac{K_{b}^{\prime}}{K_{b}}-2\frac{K_{b-1}^{\prime}}{K_{b-1}}\right)\nonumber \\
 & \quad +2\delta^{2}\frac{K_{b}}{K_{b-1}}\left[-\frac{b\left(b+1\right)}{2\sqrt{c_{0}}}+\left(\frac{b+1}{2}\right)\frac{K_{b+1}}{K_{b}}+\frac{b\left(b-1\right)}{2\sqrt{c_{0}}}-\frac{b}{2}\frac{K_{b}}{K_{b-1}}\right].
\end{align}
The argument of the tangent function is determined from 
\begin{align}
q\delta x_{0} &= \sqrt{\left(2\kappa+1\right)\delta^{2}-\left(\delta^{4}+\alpha\right)} \nonumber\\
 & = \sqrt{\left|\alpha\right|-4c_{0}+2c_{1}\delta^{2}} \nonumber\\
 & \approx\sqrt{\left|\alpha\right|-4c_{0}}+\frac{c_{1}\delta^{2}}{\sqrt{\left|\alpha\right|-4c_{0}}}.
\end{align}
Therefore, the left-hand side of Eq.~\eqref{eq:EVCEven4} is
\begin{align}
q\delta x_{0}\tan\left(q\delta x_{0}\right)+\nu &= \sqrt{\left|\alpha\right|-4c_{0}}\tan\left(\sqrt{\left|\alpha\right|-4c_{0}}\right)+\nu\nonumber\\
&\quad+\frac{c_{1}\delta^{2}}{\sqrt{\left|\alpha\right|-4c_{0}}}\left[\tan\left(\sqrt{\left|\alpha\right|-4c_{0}}\right)+\sqrt{\left|\alpha\right|-4c_{0}}\sec^{2}\left(\sqrt{\left|\alpha\right|-4c_{0}}\right)\right].
\end{align}
Solving Eq.~\eqref{eq:EVCEven4} to $O\left(\delta^{0}\right)$ gives the following
equation 
\begin{equation}
2\sqrt{c_{0}}\frac{K_{b}}{K_{b-1}}=\sqrt{\left|\alpha\right|-4c_{0}}\tan\left(\sqrt{\left|\alpha\right|-4c_{0}}\right)+\nu.
\end{equation}
Rearranging this expression gives the result in Eq.~\eqref{eq:C0Eqn} of the main text. 

Solving Eq.~\eqref{eq:EVCEven4} to $O\left(\delta^{2}\right)$ gives 
\begin{align}
\label{eq:Xc1}
& c_{1}\left\{ \frac{1}{\sqrt{\left|\alpha\right|-4c_{0}}}\left[\tan\left(\sqrt{\left|\alpha\right|-4c_{0}}\right)+\sqrt{\left|\alpha\right|-4c_{0}}\sec^{2}\left(\sqrt{\left|\alpha\right|-4c_{0}}\right)\right]+\frac{1}{2}\frac{K_{b}}{K_{b-1}}\left(\sqrt{c_{0}}+2\frac{K_{b}^{\prime}}{K_{b}}-2\frac{K_{b-1}^{\prime}}{K_{b-1}}\right)\right\}  \nonumber\\ 
& = -1+\frac{b}{2}\frac{K_{b}}{K_{b-1}}\left(\frac{1}{\sqrt{c_{0}}}+2\frac{K_{b}^{\prime}}{K_{b}}-2\frac{K_{b-1}^{\prime}}{K_{b-1}}\right)\nonumber \\
 & \quad +2\frac{K_{b}}{K_{b-1}}\left[-\frac{b\left(b+1\right)}{2\sqrt{c_{0}}}+\left(\frac{b+1}{2}\right)\frac{K_{b+1}}{K_{b}}+\frac{b\left(b-1\right)}{2\sqrt{c_{0}}}-\frac{b}{2}\frac{K_{b}}{K_{b-1}}\right].
\end{align}
Let the expression on the left-hand side in the square brackets be denoted by $X$. 
Now we simplify the expression on the right-hand side. To do this,
we use the second and fourth relations in Eqs.~(9.6.26) of Ref.~\cite{AbramowitzStegun}, with $z=2\sqrt{c_{0}}$:
\begin{align}
K_{\lambda}^{\prime}\left(z\right) & = -K_{\lambda+1}\left(z\right)+\frac{\lambda}{z}K_{\lambda}\left(z\right).\\
K_{\lambda}^{\prime}\left(z\right) & = -K_{\lambda-1}\left(z\right)-\frac{\lambda}{z}K_{\lambda}\left(z\right).
\end{align}
Thus, we have 
\begin{align}
\frac{K_{b}^{\prime}}{K_{b}} & =  -\frac{K_{b-1}}{K_{b}}-\frac{b}{z}.\\
\frac{K_{b-1}^{\prime}}{K_{b-1}} & =  -\frac{K_{b}}{K_{b-1}}+\frac{b-1}{z}.\\
\frac{K_{b+1}}{K_{b}} & =  2\frac{b}{z}+\frac{K_{b-1}}{K_{b}}.
\end{align}
Using these identities, Eq.~\eqref{eq:Xc1} becomes
\begin{align}
Xc_{1} & = -1+\frac{K_{b}}{K_{b-1}}\left(\frac{b}{2\sqrt{c_{0}}}+b\frac{K_{b}^{\prime}}{K_{b}}-b\frac{K_{b-1}^{\prime}}{K_{b-1}}\right)\nonumber \\
 &   +\frac{K_{b}}{K_{b-1}}\left[-\frac{b\left(b+1\right)}{\sqrt{c_{0}}}+\left(b+1\right)\frac{K_{b+1}}{K_{b}}+\frac{b\left(b-1\right)}{\sqrt{c_{0}}}-b\frac{K_{b}}{K_{b-1}}\right]\nonumber \\
 & =  -1+\frac{K_{b}}{K_{b-1}}\left[\frac{b}{2\sqrt{c_{0}}}+b\left(-\frac{K_{b-1}}{K_{b}}-\frac{b}{2\sqrt{c_{0}}}\right)-b\left(-\frac{K_{b}}{K_{b-1}}+\frac{b-1}{2\sqrt{c_{0}}}\right)\right.\nonumber \\
 &   \left.-\frac{b\left(b+1\right)}{\sqrt{c_{0}}}+\left(b+1\right)\left(\frac{b}{\sqrt{c_{0}}}+\frac{K_{b-1}}{K_{b}}\right)+\frac{b\left(b-1\right)}{\sqrt{c_{0}}}-b\frac{K_{b}}{K_{b-1}}\right]\nonumber \\
  & =  0.
\end{align}
This gives the result in Eq.~\eqref{eq:C1Eqn} of the main text. 

\section{Derivation of the closed-form expression for $c_{0}$}
\label{App:C0Sol}

The order $\lambda$ modified Bessel function of the second kind is defined in Eq.~(9.6.2) of Ref.~\cite{AbramowitzStegun}:
\begin{equation}
K_{\lambda}\left(z\right)=\frac{\pi}{2\sin\left(\lambda\pi\right)}\left[I_{-\lambda}\left(z\right)-I_{\lambda}\left(z\right)\right],
\label{eq:KBessel}
\end{equation}
where the order $\lambda$ modified Bessel function of the first kind is defined in Eq.~(9.6.10) of Ref.~\cite{AbramowitzStegun}:
\begin{equation}
I_{\lambda}\left(z\right)=\left(\frac{1}{2}z\right)^{\lambda}\sum_{n=0}^{\infty}\left(\frac{1}{4}z^{2}\right)^{n}\frac{1}{n!\Gamma\left(1+\lambda+n\right)}.
\end{equation}
For small arguments $0<\left|z\right|\ll\sqrt{\lambda+1}$ the asymptotic form of $I_{\lambda}(z)$ is (see Eq.~(4), pg.~16 of Ref.~\cite{Watson}):
\begin{equation}
I_{\lambda}\left(z\right)\sim\frac{1}{\Gamma\left(1+\lambda\right)}\left(\frac{z}{2}\right)^{\lambda}.
\label{eq:IBesselSeries}
\end{equation}
Using Eqs.~\eqref{eq:KBessel} and \eqref{eq:IBesselSeries}, the limiting form of $K_{\lambda}\left(z\right)$ is
\begin{align}
K_{\lambda}\left(z\right) & = \frac{\pi}{2\sin\left(\lambda\pi\right)}\left[\frac{1}{\Gamma\left(1-\lambda\right)}\left(\frac{z}{2}\right)^{-\lambda}-\frac{1}{\Gamma\left(1+\lambda\right)}\left(\frac{z}{2}\right)^{\lambda}\right]\nonumber \\
 & = \frac{1}{2}\left[\Gamma\left(\lambda\right)\left(\frac{2}{z}\right)^{\lambda}-\frac{\Gamma\left(1-\lambda\right)}{\lambda}\left(\frac{z}{2}\right)^{\lambda}\right].
\end{align}
In order to apply this result to Eq.~\eqref{eq:C0Eqn}, we require $2\sqrt{c_{0}}\ll\sqrt{\nu+\frac{1}{2}}$. In terms of $\alpha$, this is equivalent to the condition $4c_{0}\ll1+\sqrt{\frac{1}{4}+\alpha}$. 
Assuming this condition is satisfied, Eq.~\eqref{eq:C0Eqn} can be simplified to
\begin{equation}
c_{0}\approx\frac{1}{4}\left\{ \left[\sqrt{\left|\alpha\right|-4c_{0}}\tan\left(\sqrt{\left|\alpha\right|-4c_{0}}\right)+\nu\right]\frac{\left[\Gamma\left(\nu-\frac{1}{2}\right)\left(\frac{1}{\sqrt{c_{0}}}\right)^{\nu-\frac{1}{2}}-\frac{\Gamma\left(1-\left(\nu-\frac{1}{2}\right)\right)}{\nu-\frac{1}{2}}\left(\sqrt{c_{0}}\right)^{\nu-\frac{1}{2}}\right]}{\left[\Gamma\left(\nu+\frac{1}{2}\right)\left(\frac{1}{\sqrt{c_{0}}}\right)^{\nu+\frac{1}{2}}-\frac{\Gamma\left(1-\left(\nu+\frac{1}{2}\right)\right)}{\nu+\frac{1}{2}}\left(\sqrt{c_{0}}\right)^{\nu+\frac{1}{2}}\right]}\right\} ^{2}.
\end{equation}
Taking the square root of this equation and simplifying then gives
\begin{equation}
2\frac{\Gamma\left(\nu+\frac{1}{2}\right)}{\Gamma\left(\nu-\frac{1}{2}\right)}\approx\left[\sqrt{\left|\alpha\right|-4c_{0}}\tan\left(\sqrt{\left|\alpha\right|-4c_{0}}\right)+\nu\right]\left[1-\frac{\Gamma\left(\frac{3}{2}-\nu\right)}{\Gamma\left(\frac{1}{2}+\nu\right)}c_{0}^{\nu-\frac{1}{2}}\right].
\end{equation}
We suppose that $c_{0}\ll|\alpha|$ and drop the $c_{0}$ terms appearing in the square roots in the equation above:
\begin{equation}
2\frac{\Gamma\left(\nu+\frac{1}{2}\right)}{\Gamma\left(\nu-\frac{1}{2}\right)}\approx\left(\sqrt{\left|\alpha\right|}\tan\sqrt{\left|\alpha\right|}+\nu\right)\left[1-\frac{\Gamma\left(\frac{3}{2}-\nu\right)}{\Gamma\left(\frac{1}{2}+\nu\right)}c_{0}^{\nu-\frac{1}{2}}\right].
\end{equation}
Now solve this equation for $c_{0}$ to obtain 
\begin{equation}
c_{0}\approx\left\{ \frac{\Gamma\left(\frac{1}{2}+\nu\right)}{\Gamma\left(\frac{3}{2}-\nu\right)}\left[1-\frac{2}{\left(\sqrt{\left|\alpha\right|}\tan\sqrt{\left|\alpha\right|}+\nu\right)}\frac{\Gamma\left(\nu+\frac{1}{2}\right)}{\Gamma\left(\nu-\frac{1}{2}\right)}\right]\right\} ^{\frac{1}{\nu-\frac{1}{2}}}.
\end{equation}
After inserting the expression for $\nu$ from Eq.~\eqref{eq:Nu}, the result in Eq.~\eqref{eq:C0Eqn2} of the main text is obtained. 
\end{widetext}

\bibliographystyle{apsrev4-1}
\bibliography{References}

\end{document}